\documentclass{article}

\usepackage{PRIMEarxiv}

\usepackage[utf8]{inputenc} % allow utf-8 input
\usepackage[T1]{fontenc}    % use 8-bit T1 fonts
\usepackage{hyperref}       % hyperlinks
\usepackage{url}            % simple URL typesetting
\usepackage{booktabs}       % professional-quality tables
\usepackage{amsfonts}       % blackboard math symbols
\usepackage{nicefrac}       % compact symbols for 1/2, etc.
\usepackage{microtype}      % microtypography
\usepackage{lipsum}
\usepackage{fancyhdr}       % header
\usepackage{graphicx}       % graphics
\usepackage{amsmath,amsthm,amsfonts,amssymb}
\usepackage{graphicx,cite,enumerate}
\usepackage[table,xcdraw]{xcolor}
\usepackage{float}
\usepackage{tikz}
\def\checkmark{\tikz\fill[scale=0.4](0,.35) -- (.25,0) -- (1,.7) -- (.25,.15) -- cycle;} 
\usepackage{float}
\usepackage{subfig}

\graphicspath{{media/}}     % organize your images and other figures under media/ folder

\title{UAS in the Airspace: A Review on Integration, Simulation, Optimization, and Open Challenges
}

\author{
  Euclides C. P. Neto, Derick M. Baum, Jorge Rady de Almeida Jr., João B. Camargo Jr., Paulo S. Cugnasca \\
  Safety Analysis Group - School of Engineering (POLI) \\
  University of São Paulo (USP) \\
  São Paulo, Brazil\\
  \texttt{\{euclidescpn, derick.baum, jorgerady, joaocamargo, cugnasca\}@usp.br} \\
}

\begin{document}
\maketitle

\begin{abstract}
Air transportation is essential for society, and it is increasing gradually due to its importance. To improve the airspace operation, new technologies are under development, such as Unmanned Aircraft Systems (UAS). In fact, in the past few years, there has been a growth in UAS numbers in segregated airspace. However, there is an interest in integrating these aircraft into the National Airspace System (NAS). The UAS is vital to different industries due to its advantages brought to the airspace (e.g., efficiency). Conversely, the relationship between UAS and Air Traffic Control (ATC) needs to be well-defined due to the impacts on ATC capacity these aircraft may present. Throughout the years, this impact may be lower than it is nowadays because the current lack of familiarity in this relationship contributes to higher workload levels. Thereupon, the primary goal of this research is to present a comprehensive review of the advancements in the integration of UAS in the National Airspace System (NAS) from different perspectives. We consider the challenges regarding simulation, final approach, and optimization of problems related to the interoperability of such systems in the airspace. Finally, we identify several open challenges in the field based on the existing state-of-the-art proposals.
\end{abstract}

% keywords can be removed

\section{Introduction}
% Airspace operation and challenges
\par Air transportation is essential for society, and it is increasing gradually due to its importance \cite{marquart2003future} \cite{6317129}. The growth in flights number makes the airspace more complex while leading to higher revenue. There are several obstacles to be overcome by authorities in the following years in terms of safety and efficiency of airspace. The Air Traffic Control (ATC) is pivotal in optimizing airspace, assuming that safety and efficiency are key aspects of airspace operation \cite{girdner2016integrated} \cite{7898410} \cite{526911}. The ATC is divided into ATC units, which are ``generic term meaning variously, area control center, approach control unit or aerodrome control tower" \cite{ICAO2016}. These units are arranged to accommodate all airspace users by creating sectors. The role of controlling aircraft in each control sector is played by Air Traffic Controllers (ATCo), who communicate to ATCos responsible for other sectors to provide smooth conduction of aircraft throughout their flights.

% Air Traffic Controller
\par The ATC targets offering suitable levels of safety and efficiency and addressing complex situations. Moreover, ATC provides Air Traffic Services (ATS) to flights through ATCo instructions. The primary objective of these services includes avoiding mid-air collisions and collisions with obstructions and optimizing and maintaining the flow of air traffic \cite{ICAO2001}. The ATCo conducts the aircraft in a sector by applying techniques to improve safety and efficiency (e.g., vectoring). These professionals act collaboratively from the beginning to the end of each flight, and other ATCos are assigned to control such flights once a new sector is reached. Conversely, an obstacle currently faced is to maintain workload\footnote{Workload can be defined as a metric that represents the difficulty of ATCo in understanding a particular situation \cite{meckiff1998tactical} and can be expressed in terms of seconds.} level under an acceptable threshold.

% ATCo Workload
\par Among the several safety threats in airspace operation, mid-air collision can be highlighted, which depends on a set of events despite issues in aircraft mechanical systems, such as high ATCo workload levels and loss of the established minimum separation. There is an effort of authorities toward such events (e.g., ATCo training for critical situations and design of safe standard procedures). Furthermore, in cases of high air traffic density, a safer measure of the capacity of a sector is based on ATCo workload \cite{majumdar2001estimating} \cite{pejovic2020relationship}, i.e., the number of aircraft that can be safely accommodated decreases when there is a higher workload level. As ATCo workload levels are related to safety and there is an understanding by research and operational community that airspace complexity is one of the main factors that impact this metric \cite{majumdar2002factors}, situations that these professionals are not familiar with tends to be more unsafe. Moreover, several variables compose complexity, such as traffic density and mental factors \cite{dervic2015atc}.

% Technological development of airspace
\par To improve the airspace operation, new technologies are under development, such as Unmanned Aircraft Systems (UAS) \cite{austin2011unmanned} and Decision Support Tools (DST) for ATCos (e.g., Arrival and Departure managers) \cite{noskievivc2017air}. These new technologies present advantages in many aspects, such as safety, efficiency, and airspace capacity. Furthermore, the DSTs aim to lead ATCos to more effective decisions, which tends to reduce the ATCos workload and, ultimately, to reduce airspace complexity \cite{majumdar2002factors}. Although these technologies are used in different situations, they may bring uncertainties since it is reasonable to consider that ATCos may not be familiar with them. Furthermore, new technologies being integrated into the airspace nowadays (e.g., UAS) may be typical in the future, increasing this familiarity.

% Unmanned Aircraft System (UAS) Overview
\par Moreover, the UAS is vital to different industries due to its advantages brought to the airspace (e.g., efficiency) \cite{Ganti2016Implementation}. The UAS has been considered a relevant topic in the engineering community due to its applications \cite{fasano2016Sense} and consists of systems composed of subsystems such as Unmanned Aerial Vehicle (UAV), its payloads, the control station, and communications subsystems \cite{austin2011unmanned} \cite{fasano2016Sense}. Different types of UAS (e.g., Autonomous Aircraft - AA - and Remotely Piloted Aircraft Systems - RPAS) present different subsystem requirements (e.g., remote piloting interfaces are present in RPAS but not in manned aircraft). For instance, the ground station used to pilot RPASs is not part of the AAs, which are considered fully autonomous.

% UAS Integration
\par In the past few years, there has been a growth in UAS numbers \cite{guerin2015consideration} in segregated airspace. However, there is an interest in integrating these aircraft into the National Airspace System (NAS). These aircraft, which have several military and civil applications, present challenges to their integration to be faced by authorities in terms of safety, i.e., new ways of reaching unsafe states are included in the airspace. For instance, bugs in software may maneuver the aircraft and lead it to undesired headings. Also, considering RPAS, failures in Command and Control (C2) link, i.e., the connection pilots use to communicate to the aircraft, may lead to unsafe states \cite{neto2017airspace} \cite{ICAO2015}.

% ATCo and UAS
\par The relationship between UAS and ATC needs to be well-defined due to the impacts on ATC capacity these aircraft may present. Throughout the years, this impact may be lower than it is nowadays because the present lack of familiarity in the relationship between UAS and ATCo contributes to higher workload levels. As UAS only operate in segregated airspaces, ATC tends to be more concerned when controlling a gate-to-gate flight of these autonomous systems. Different challenges to enable this integration must be addressed, such as specific regulations, policies, and procedures, enabling technologies and standards development for dealing with UAS \cite{grindle2016unmanned}. As the integration of UAS enables new applications and its use may increase in the future \cite{gupta2013review}, developing approaches to integrate it safely is essential.

% TMA
\par Furthermore, the Terminal Control Area (TMA) is a critical control area generally established at the confluence of Air Traffic Service (ATS) routes in the vicinity of one or more major aerodromes \cite{ICAO2001}. In this area, the aircraft tend to be closer to each other. In general, TMA is the most resource-constrained component of the air transportation system due to the number of aircraft that operate simultaneously \cite{khadilkar2016integrated}. Its complexity increases according to the airspace configuration (e.g., traffic density and weather conditions). Hence, operations in the TMA usually follow standard procedures established, e.g., Standard Instrument Departure (SID) and Standard Instrument Arrival (STAR).

% Final approach
\par However, standard procedures (e.g., STAR) cannot be followed in some cases, e.g., in high traffic density. In these cases, a highly challenging ATCo task is the sequencing of the aircraft during the approach, considering the arrival segment and the final approach \cite{ICAO2006} \cite{ahmed2017evolutionary} \cite{9508434} due to complex maneuvers constraints. To accomplish this, the aircraft are conducted in a way to avoid conflict, i.e., in a way not to disregard the minimum separation requirements, as well as to avoid flying through cumulonimbus (CB), which are cloud formations that present a real impact on aviation \cite{fromm2005pyro}. Finally, the primary objective of defining a final arrival segment is to deliver a set of aircraft from the final sector of the TMA to the final phase of its landing procedure (i.e., the final approach), taking operational efficiency and safety into consideration.

% Problem in optimizing safety and efficiency during final approaches
\par Establishing final arrival segments for achieving optimized aircraft operation in terms of safety and efficiency is not a simple task. From the safety perspective, the ATCo workload related to conflict avoidance during this phase, i.e., aircraft minimum separation from other aircraft and to the cumulonimbus (CB), must remain at acceptable levels once an increase in this metric may present impacts on safety levels. From an efficiency perspective, the aircraft set must be delivered to the airport as soon as possible. Depending on the scenario, the ATCo must act rapidly to avoid airspace reaching unsafe states. As the number of aircraft increases, the situation becomes more complex and, consequently, more difficult to be controlled by the ATCo.

% Problem of UAS integration
\par On the other hand, integrating UAS in the NAS airspace is a challenge nowadays. According to ICAO \cite{ICAO2005}, ``the airspace will be organized and managed in a manner that will accommodate all current and potential new uses of airspace, for example, Unmanned Aerial Vehicles (UAV) and that any restriction on the use of any particular volume of airspace will be considered transitory". Furthermore, although rules for UAS flights are defined for segregated airspace \cite{ICAO2015}, the increasing interest in the usage of UAS for different applications (military and civilian) leads to a need for integrating them into the National Airspace System (NAS). To accomplish this, safety levels must not be compromised \cite{ICAO2015}. 

% Problem of UAS integration into final approach
\par Toward the challenges faced in the final sector in complex situations, the presence of UAS is an important player. Due to lack of familiarity, it is reasonable to consider that the ATCo may feel uncomfortable in controlling autonomous aircraft, which is a result of the uncertainty on UAS operation \cite{zlotowski2017can} \cite{8453455}. However, the arrival procedure is a critical and complex task even without the UAS presence, and definition sequencing solutions for both Manned Aircraft (MA) and UAS, especially during the early stages of UAS integration in the National Airspace System (NAS), may lead to higher ATCo workload levels. Furthermore, there needs to be more simulation methods that include the UAS in the final sector and include complex situations (e.g., bad weather conditions).

% Problem of not measuring the familiarity of ATCos with UAS
\par Finally, measuring the familiarity of ATCo with a particular aircraft (e.g., UAS) is a challenge. Not only because UAS does not operate in the NAS nowadays, but also because of the relationship between familiarity and cognition. Measuring familiarity enables better sequencing solutions in arrival procedures, especially from the ATCo workload perspective, i.e., from the safety perspective.

\par
\begin{figure}[H]
\centering
\includegraphics[width=1\linewidth]{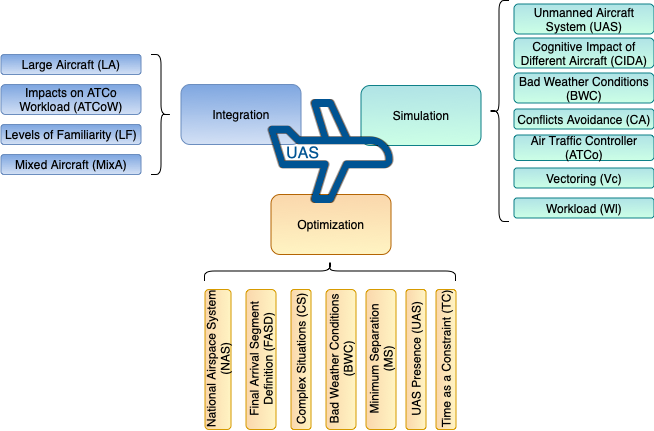}
\caption{Main aspects analyzed for UAS Integration, Simulation, and Optimization.}
\label{fig:aspects}
\end{figure}

 Thereupon, the main goal of this research is to present a comprehensive review of the advancements in the integration of Unmanned Aircraft Systems (UAS) in the National Airspace System (NAS) from different perspectives. We consider the challenges regarding simulation, the final approach, and optimization of problems related to the interoperability of such systems in the airspace. We also highlight several open challenges in the field based on the existing state-of-the-art proposals. Finally, Figure \ref{fig:aspects} illustrates the main aspects analyzed for UAS Integration, Simulation, and Optimization. For each area, several aspects are taken into consideration based on their relevance in the reviewed works. Finally, some aspects are included in more than one area.

This article is organized as follows: Section \ref{sec:Integration} reviews strategies focused on the UAS Integration in the National Airspace System (NAS). Sections \ref{sec:simulation} and \ref{sec:arrival} analyze airspace simulation and arrival segment optimization efforts considering the UAS presence. Finally, Sections \ref{sec:open_challenges} and \ref{sec:conclusion} present the open challenges and conclusions of this research, respectively.

\section{UAS Integration in the National Airspace System (NAS)}
\label{sec:Integration}
\par This section presents works related to approaches of including and measuring impacts of Unmanned Aircraft Systems (UAS) integration in the airspace from different perspectives. The works presented in this section are classified according to the presence of the critical aspects: Large Aircraft (LA), Impacts on ATCo Workload (ATCoW), Levels of Familiarity (LF), and mixed aircraft (MixA - UAS and Manned Aircraft operating together).

% paper number 1
% Considered Large Aircraft - x
% Considered Evaluation of ATCo Workload - x
% Considered Different levels of Familiarity - x
% Considered Mixed Aircraft (UAS and Traditional Aircraft) - √
\par Shmelova et al. \cite{Shmelova2016modelling} present an approach based on statistical data to deal with the problem of Unmanned Aerial Vehicles (UAV) flights considering different tasks in emergencies, which are special situations and tend to increase the Air Traffic Controller (ATCo) workload. Also, an analysis of the emergency type is conducted and a sequence of actions is defined. The authors present a motivation for the development of their research, which includes the lack of algorithms to recommend actions for the UAV operator in an emergency, problems in the decomposition of the decision-making process and the lack of structure of Distributed Decision Support System (DDSS), which aims to recommend actions to appropriate aircraft from a global perspective, for remotely piloted aircraft. Furthermore, models are developed by the authors to determine the optimal landing site in specific situations and search for optimal flight routes. However, this effort only considers emergencies, and the proposed model does not consider complex airspace, i.e., airspace with many aircraft. Finally, impacts on ATCo workload due to UAS presence and lack of familiarity of ATCo with this new technology are not taken into account.

% paper number 2
% Considered Large Aircraft - x
% Considered Evaluation of ATCo Workload - √
% Considered Different levels of Familiarity - √
% Considered Mixed Aircraft (UAS and Traditional Aircraft) complex - x
\par Pastor et al. \cite{pastor2014real} aim to evaluate the interaction between a Remotely Piloted Aircraft System (RPAS) and Air Traffic Management (ATM) considering that an RPAS is being operated in shared airspace, i.e., along with traditional aircraft in National Airspace System (NAS). This evaluation employs human-in-the-loop real-time simulations that allow simulating activities from the RPAS Pilot-in-Command (PiC) and the Air Traffic Controller (ATCo), and three different perspectives: the separation management, the contingency management, and the capacity impact in the overall ATM system. The experiments conducted, which were realistic and without excessive complexity, presented recommendations to improve the evaluation, e.g., preliminary analysis of traffic to prevent separation conflicts and improvement of ADS-C flight intent communication mechanism. However, this research does not consider complex airspace scenarios regarding the number of aircraft.

\par The authors in \cite{allignol2016integration} propose a geometrical horizontal detect and avoid algorithm for UASs operation, based on ADS-B-equipped aircraft information, in Terminal Control Areas (TMA), considering a constant speed. This approach employed recorded commercial traffic trajectories and random conflict scenarios with UASs. The main goal is to show the algorithm's applicability in ensuring the separation from traditional aviation, i.e., this research considers a mix of manned and unmanned aircraft. Also, six different missions are considered, such as flying straight or turning and climbing or descending. Other important aspects observed were the influence of the various parameters on the separation achieved and the number of maneuvers required, i.e., the strategy used selects the best directions respecting the range of heading degrees allowed. The experiments showed the proposal's effectiveness, which maintains the heading constant and changes it robustly if the minimum separation threshold is greater than the current separation between the UAS and a given aircraft. One should note that these methods were tested on 2850 realistic traffic scenarios, which were issued from data recorded in a French Terminal Control Area (TMA). However, although there is a considerable effort in the detection and avoidance process, the authors do not consider UAS a large aircraft. Finally, the ATCo workload (e.g., the additional cognitive workload related to UAS operation) is not evaluated.

% paper number 5
% Considered Large Aircraft - √
% Considered Evaluation of ATCo Workload - x
% Considered Different levels of Familiarity -x
% Considered Mixed Aircraft (UAS and Traditional Aircraft) - √
% STEPWISE INTEGRATION OF UAS IN NON-SEGREGATED AIRSPACE – THE POTENTIAL OF TAILORED UAS ATM PROCEDURES
\par The authors in \cite{korn2012stepwise} focus on possible guidelines for UAS integration in the National Airspace System (NAS). The main objective of this approach is to maintain the level of safety of UAS and traditional aircraft nearly the same, which may lead the authorities to implement new airspace rules such as additional separation for unmanned traffic. The authors also consider the usage of Airborne Collision Avoidance System (ACAS) maneuvers and avoidance logic. In this work, the authors conducted the experiments considering a series of simulations that present a reduction in conflict potential (UAS and traditional traffic). The reduction of impact on airspace operation, considering the UAS integration, is also highlighted since its integration in the NAS is a challenge in terms of future acceptance of these autonomous systems. In this context, hazardous situations related to UAS operation are stated, such as UA leaving cleared planned route, ATC having no position information, and loss of communication. Furthermore, the reader should note that UAS flights can be conducted with low interference considering proper mission planning, although ATC needs to control these autonomous systems, which tends to increase the workload of the ATCos. The authors also suggest the presence of specialized UAS controllers, which could share the duty of controlling the airspace among many ATCos. However, although the workload is a concern of this paper, a workload evaluation is not conducted. Also, the workload related to the operation of UAS does not include the additional cognitive workload present, especially in the early stages of UAS integration. Also, different levels of familiarity of ATCo with these
systems are not considered.

% paper number 6
% Considered Large Aircraft - √
% Considered Evaluation of ATCo Workload - x
% Considered Different levels of Familiarity - x
% Considered Mixed Aircraft (UAS and Traditional Aircraft) - √ ????
\par An approach for safety and risk assessment in unmanned aerial system (UAS) operations, which employs stochastic fast-time simulation and is conducted in the NAS, is presented in \cite{de2016approach}. Considering that the integration of UAS in the NAS is a concern to airspace regulators, the main goal of this research is to calculate fatality risk and to measure how different aspects and phases of UAS operations increase the risk. To accomplish this, the authors model and simulate UAS operations over a specific hazardous region by applying different stochastic parameters (e.g., altitude, performance variation, and latency). Note that the risk analysis considers fatalities and is based on published ground impact models, which enable the usage of fast-time simulation to assess specific situations. Furthermore, the method adopted in this research, which compared different risk analysis models, is important to highlight mitigation actions for all stakeholders in the safety assessment. However, although this paper discusses the importance of accurately measuring the risks of fatalities in UAS operations, some aspects are not considered. For instance, the workload associated with the presence of UAS in the airspace is not faced. Thus, the level of familiarity of ATCo with this technology is not considered.

% paper number 7
% Considered Large Aircraft - x
% Considered Evaluation of ATCo Workload - x
% Considered Different levels of Familiarity - x
% Considered Mixed Aircraft (UAS and Traditional Aircraft) - √
% A Design Study for the Safe Integration of Unmanned Aerial Systems into the National Airspace System
\par In \cite{branch2016design}, the effectiveness of geofencing systems (such as static and dynamic design) for UAS, which defines geographical boundaries in specific airspace areas, is analyzed. The authors also compare the geo-fencing effectiveness to the current and traditional proposed regulations on collision avoidance systems. To accomplish this, Monte Carlo simulations are employed, considering growth and incident rates based on the incident data. In this context, plenty of UAS (more than 1 million) are available to operate within the National Airspace (NAS), but there is a need to integrate them safely into the NAS. This process must be conducted to optimize the relationship between cost and safety. Furthermore, UAS is considered disruptive technology to be included in NAS, and operations cost reduction motivates such integration. Although even considering the substantial growth of these aircraft in the past few years and so forth, the step-wise increase of operational tests and global acceptance, the number of incidents has also grown. This growth has been due to different reasons, such as the disobedience of planned altitude and location by UAS. The experiments showed that UAS operations conducted into regulated thresholds, i.e., to specific geographical areas, provide a cost-effective method that respects safety levels and eliminates 98\% of the UAS incidents as reported by FAA. However, this research does not consider aspects related to ATCo operation, such as workload.

% paper number 8
% Considered Large Aircraft - √
% Considered Evaluation of ATCo Workload - x
% Considered Different levels of Familiarity - x
% Considered Mixed Aircraft (UAS and Traditional Aircraft) - √
\par Gimenes et al. \cite{ricardo2014guidelines} propose guidelines to support UAS regulations for the integration of fully autonomous UASs into the Global Air Traffic Management (GATM) System and, consequently, into shared airspace. These guidelines are proposed facing three perspectives: the aircraft itself, the Piloting Autonomous System (PAS), and the integration of autonomous UASs in the National Airspace System (NAS). Considering that there are social and economic interests in UAS applications, enabling this technology to operate along with Manned Aircraft (MA) has considerable potential. The main issue of this integration is that aeronautical authorities should regulate UAS operations in the NAS, although defining these rules is difficult since there is not a deep understanding of UAS operations and how they behave in case of failures (e.g., contingency operations). Throughout the paper, the authors present the guidelines with different focuses. For instance, regarding the ``aircraft focus", although it is not in the scope of this paper, the authors state that it ``should be submitted to at least the same processes and criteria of developing, manufacturing and certification regarding avionic systems of manned aircraft, aiming to reach the same safety levels". Furthermore, the authors highlight that the UAS concept should be based on aeronautical precepts and that the possibility of integrating UASs into airspace depends on specific regulations. However, this research does not consider the ATCo evaluation and the impact of UAS operation on ATCo performance.

% paper number 9
% Considered Large Aircraft - √
% Considered Evaluation of ATCo Workload - x
% Considered Different levels of Familiarity - x
% Considered Mixed Aircraft (UAS and Traditional Aircraft) - √

%Integration of Unmanned Aircraft System (UAS) in non-segregated airspace: A complex system of systems problem
\par In \cite{ramalingam2011integration}, the authors present a discussion on the integration of UAS in the NAS. This problem is a complex system-of-systems problem, considering the level of difficulty higher the technical challenges related to the development, implementation, and validation of this technology. Considering that the system design itself is a complex problem, the authors emphasize that the operation of UAS into NAS depends on aviation regulatory authorities, but this sort of regulation is not simple to be defined. The main challenge identified is to design UASs with high safety standards that operate, such as manned aircraft (e.g., transparency). UAS numbers have increased tremendously in the last few years due to the distinct capabilities and cost advantages compared to manned aircraft in most situations, enabling these aircraft to operate alongside manned aircraft is desirable. Throughout this paper, different regulations are presented, such as regulations followed in Australia. Furthermore, this paper analyzes reasons for the difficulty in integrating UAS in the NAS. However, although this research considers workload an important aspect of UAS inclusion, it does not propose an approach to evaluate it. Finally, the evolution in terms of the familiarity of the relationship UAS-ATCo is not considered.

% paper number 10
% Considered Large Aircraft - √
% Considered Evaluation of ATCo Workload - X
% Considered Different levels of Familiarity - X
% Considered Mixed Aircraft (UAS and Traditional Aircraft) - √
\par In \cite{kamienski2015atc}, the authors aim to identify potential ways of mitigating issues related to different UAS challenges. Also, a revision of some of the pros and cons of these different approaches and recommendations for changes in procedures, automation, and policy. The impacts of an integrated UAS operation on ATC are not fully clear yet, even in less congested areas, but there is a need to integrate this aircraft in terms of cost reduction and efficiency. The MITRE Corporation, which is the corporation of the authors of this research, has been using techniques to identify the impacts of UAS on ATC in the past years, which has shown that, for instance, the process of filing flight plans, the usage of latitude/longitude waypoints instead of named fixes or waypoints and possible delays or loss of communication have considerable impact. More specifically, the authors state that the impacts are presented in five major areas: UAS flight planning and automation, the UAS control link (delays and loss), UAS-specific information and procedures, ATC training, and UAS interaction with the future NAS. However, although this research highlights challenges of ATC in terms of UAS integration and considers ATCo workload as an impacted metric, the level of familiarity of ATCo with a specific aircraft is not considered, i.e., there is not a workload evaluation process that highlights the difference between UASs of different familiarity (from the ATCo perspective).

% paper number 11
% Considered Large Aircraft - X
% Considered Evaluation of ATCo Workload - X
% Considered Different levels of Familiarity - X
% Considered Mixed Aircraft (UAS and Traditional Aircraft) - √
\par The authors in \cite{mcfadyen2016terminal} deal with the problem of integrating UASs above urban environments, i.e., into low-altitude airspace. This integration includes major Terminal Manoeuvring Areas (TMA) and helicopter landing sites. A set of data-driven modeling techniques are employed to assess existing air traffic as starting for UAS operation. To accomplish this analysis, the authors exploit low-altitude air traffic data sets in order to discover existing no-fly zones and an alternative geometric approach to defining exclusion zones, which is applied to a real region (Australia), including one International airport and helicopter landing area. Considering that determining adequate exclusion zones for unmanned aircraft in an urban environment is an important task and that regulations may, in some cases, include UAS in these areas without considerable reduction of risks of collision, the main goal of this research is to propose an approach to define exclusion zone appropriately. The results showed a need for more rigorous scientific approaches to safely integrate these autonomous aircraft into shared and urban airspaces. However, although this work constitutes an important and unique contribution to UAS integration in the urban environment, aspects such as workload measurement during the definition of these areas are not considered.

% paper number 12
% Considered Large Aircraft - x
% Considered Evaluation of ATCo Workload - x
% Considered Different levels of Familiarity - x
% Considered Mixed Aircraft (UAS and Traditional Aircraft) - √
\par The authors in \cite{clothier2017making} propose a way to create a Risk-Informed Safety Case (RISC) applied to the context of small UAS operation safety assurance. This approach aims to facilitate safe and cost-effective operations of small UAS by presenting the comprehensive measures considered to eliminate, reduce, or control the safety risk. The RISC proposed comprises barrier models of safety, which support the development of safety measures, and structured arguments, which assure safety in operations (through, for instance, appropriate evidence). The authors also propose a model for small UAS operational risk, which considers, for instance, specific hazards (e.g., mid-air collision) and operational risks which depend on the small UAS. Ultimately, this paper shows key safety-related assurance concerns to be addressed and the development of a layered framework for reasoning about those concerns, which can be useful for regulators and various stakeholders in justifying confidence in operational safety in the context of the absence of the relevant aviation regulations for small UAS. However, although the authors focused on proposing an approach to deal with the current state, i.e., a lack of presence of UAS in shared airspace, this research does not measure the impact of these aircraft on ATCo operation (e.g., workload) and, ultimately, into safety levels. Finally, different levels of aircraft familiarity to the ATCo are not considered.

% paper number 13
% Considered Large Aircraft - √
% Considered Evaluation of ATCo Workload - X
% Considered Different levels of Familiarity - X
% Considered Mixed Aircraft (UAS and Traditional Aircraft) - √ ?????
\par In \cite{washington2017managing}, a new framework for system safety certification under conditions of uncertainty is proposed considering a Bayesian approach to the modeling of the average probability of failure conditions. Nowadays, the debates over developing appropriate system safety requirements for UAS are heated. An interesting point of view is approaching this analysis by determining the allowable average probability per flight hour of failure conditions. Due to the lack of knowledge and data to inform the assessment of failure probabilities of UAS, a level of uncertainty may be considered during the system safety assessment (SSA) process, which presents many advantages. The conducted experiments showed the suitability of the proposed approach's safety measures. Thus, other sources of uncertainty are intended to be considered in future works. Finally, the authors state that using a constant failure rate model is challenged by using a Weibull distribution, which seems to be a more appropriate representation of UAS failure occurrence. However, although there is an effort to estimate UAS failures and an interesting approach that relates uncertainty to safety assessment that can be applied to small and large UASs is presented, this research does not focus on aspects related to ATCo operation, such as communication to UAS.

% paper number 14
% Considered Large Aircraft - X
% Considered Evaluation of ATCo Workload - X
% Considered Different levels of Familiarity - X
% Considered Mixed Aircraft (UAS and Traditional Aircraft) - √
\par Romero et al. \cite{romero2017proposal} discuss on present and future of the Remotely Piloted Aircraft System (RPAS) in terms of regulation by aeronautical authorities. This discussion considers different countries (e.g., Colombia, Malta, and Japan) aiming to understand the integration of RPAs in the NAS from the ATCo perspective. An analysis of the existing classification types of RPAS (classes one, two, and three) is conducted. Moreover, the results of integrating three RPAS in the NAS, successfully performed in a real setting, from the air traffic control center in Barranquilla (Colombia) are presented. Note that there were no losses of separation with other aircraft or between RPAS and that one of the authors of this paper, an air traffic controller of Barranquilla, coordinated the different entities that participated in the implementation of this successful operation of integrating RPA in the NAS. Finally, a proposal is made to integrate this type of aircraft in the NAS, which considers airspaces classification, RPA classification (in terms of navigation performance), and contingency operation. However, although this paper is an outstanding contribution due to the integration of RPAS into shared airspace in a real setting, the authors do not consider future scenarios in which RPAS may be represented by commercial aircraft. Finally, different types of aircraft in terms of ATCo familiarity are not considered.

% paper number 15
% Considered Large Aircraft - X
% Considered Evaluation of ATCo Workload - √
% Considered Different levels of Familiarity - X
% Considered Mixed Aircraft (UAS and Traditional Aircraft) - √
\par The basis to implement a risk model and general methodologies to investigate RPAS safety, according to the operational scenarios defined by European Aviation Safety Agency (EASA), is proposed in \cite{grimaccia2017risk}. The authors analyzed results achieved in experimental flights of multiple RPAS. As the modern aeronautical scenario is being adapted to accommodate new key elements, including the Remote Piloted Aircraft Systems (RPAS), initially used for military purposes only, this new sort of aircraft is ready to become a new airspace user in civilian applications and, even considering that it cannot operate in the NAS nowadays, there is a potential growth expectation in terms of investments on this technology. This research points out the hazards related to RPAS operation in the NAS, such as failures in Command and Control (C2) link, ATCo performance referred to with high workload situations, pilots' performance with high workload situations, external factors (e.g., emergencies), and jamming. Moreover, the authors highlight that a requirement for disclosing the airspace to RPAS is the implementation of a specific Safety Management System (SMS) for every aeronautical operator. Finally, the preliminary risk analysis presented in this research highlights many possibilities to be further investigated in future works. However, although this approach can be easily extended from small to large RPAS, this research does not focus on the different maturity levels that each aircraft may present in the relationship with the ATCo, which may have a considerable impact on workload.

% paper number 16
% Considered Large Aircraft - √
% Considered Evaluation of ATCo Workload - X
% Considered Different levels of Familiarity - X
% Considered Mixed Aircraft (UAS and Traditional Aircraft) - √
\par Perrottet \cite{perrottet2017enabling} explores the challenges related to the application of Performance Based Navigation (PBN) in UAS operation, which include GNSS navigation, layered PBN procedures to UAS performance characteristics, and the capability of performing instrument procedures (in case of failures in communication link). The main goal of this integration is to enable UAS to fly without limitations in airspace shared with other aircraft. However, the primary focus of integrating these aircraft has been identifying a way to compensate for the lack of a human pilot onboard, such as Detect and Avoid (DAA) and Datalink technologies. The author also states that safety and efficiency are two key metrics of airspace and that they may or may not be inherently linked as in manned aviation, i.e., UASs may provide a more independent relationship between safety and efficiency for specific operations. Finally, this new balance between safety and efficiency must aim to maintain the high level of safety observed in today’s NAS, which is a requirement to turn the advantages provided by UAS reasonable. However, although the authors deal with the problem of integrating UAS into shared airspace, this research presents an overview of the challenges faced. One should note that large aircraft are also considered, but aspects such as ATCo workload are not considered.

% paper number 17
% Considered Large Aircraft - √
% Considered Evaluation of ATCo Workload - X
% Considered Different levels of Familiarity - X
% Considered Mixed Aircraft (UAS and Traditional Aircraft) - √
\par In \cite{sesso2016approach}, the authors propose a qualitative approach for assessing the safety of UAS operations when using Automatic Dependent Surveillance-Broadcast (ADS-B) systems considering a new testing platform, which is called PIpE-SEC, as a possible approach for this safety evaluation. This research focuses on the influence of data integrity, which is considered a safety-related parameter. The increase in UAS numbers is pressing authorities to design airspace rules to integrate safely, although safety issues arise when both manned and unmanned aircraft coexist in the airspace. Furthermore, surveillance and data integrity play important roles in controlling these aircraft. In this context, the positional information provided by the ADS-B, which is essential to UASs control systems operation, interacts with the Sense and Avoid Systems (S\&AS) of the UAS to avoid exposure to unsafe situations. Finally, the authors discuss the usage of a methodology previously applied to manned systems for assessing safety and state that the adoption of the presented methodology and tools enables the identification of appropriate scenarios for the insertion of UAS along with manned aircraft, maintaining the same safety. However, this research does not consider the impacts of positional errors on aircraft with different maturity levels. For instance, the impacts of positional error of UAS in the early stages of its integration and in the long-term stage are not considered.

% paper number 18
% Considered Large Aircraft - √
% Considered Evaluation of ATCo Workload - X
% Considered Different levels of Familiarity - X 
% Considered Mixed Aircraft (UAS and Traditional Aircraft) - √
\par Oztekin et al. \cite{oztekin2012development} propose a systems-level approach to analyze the safety impact based on risk controls of introducing new technologies into the NAS, such as UAS, considering Safety Management Systems (SMS) principles and existing regulatory structure. Furthermore, the authors present a methodology to identify minimum safety baselines for safe operations in the NAS and show its applicability through a proof-of-concept study. In this context, UAS emerges as a viable technology for potential civil and commercial applications in the NAS, although it brings the need for a deeper analysis of safety impact. A detailed outline of the concepts and methodologies used for constructing a proof-of-concept study for the proposed approach, which considers related hazards and underlying causal factors, is also presented. Finally, the safety baseline proposed in this research identifies a set of minimum risk controls for conducting safe operations. In future steps, the authors intend to focus on identifying the UAS-specific components of the developed safety baseline to identify hazards related specifically to the UAS domain. However, although this research considers scenarios with both manned and unmanned aircraft, aspects such as ATCo workload are not considered.

% paper number 19
% Considered Large Aircraft - √
% Considered Evaluation of ATCo Workload - X
% Considered Different levels of Familiarity - X
% Considered Mixed Aircraft (UAS and Traditional Aircraft) - √
\par An architecture that provides data and software services to enable a set of UAS platforms to operate in the NAS (including, for instance, terminal, en route, and oceanic areas) is presented in \cite{heisey2013reference}. The authors present the general architecture and a Sense and Avoid (SAA) testbed implementation to quantify the benefits. This architecture, based on a Service Oriented Architecture (SOA) with open standards, aims to support UAS operations by offering services to meet their specific requirements, such as command, control, and data management. This proposed approach is considered guidance and offers architectural best practices. Finally, even considering that an SOA architecture makes some aspects of certification more challenging, this approach presents some advantages and can be implemented to meet performance requirements. One should note that certification may be more straightforward considering the usage of formal service contracts, comprehensive interface and quality of service specifications, and governance process in this SOA architecture. However, this research does not provide specific services considering each aircraft's different maturity levels. Also, although this contribution focuses on integrating UAS in the NAS, aspects such as impacts on ATCo workload are not considered.

% paper number 20
% Considered Large Aircraft - X
% Considered Evaluation of ATCo Workload - X
% Considered Different levels of Familiarity - X
% Considered Mixed Aircraft (UAS and Traditional Aircraft) - √
\par Wargo et al. \cite{wargo2017enhancing} presents an integrated view of how enabling technologies can support the Remote Pilot in Command (PIC) and the UAS operations in congested terminal airspace operations. There is a desire, nowadays, to integrate large and small UAS (e.g., RPAS) into the complex terminal environment and the airport surface. The new surveillance technologies that are under development, as well as the access to the NAS system information via System Wide Information Management (SWIM), are manners for improving the remote UAS Pilot in Command’s (PICs) performance and, consequently, to conduct UAS operations safely in the terminal environment. Vendors can get data feeds for, for instance, flight planning, airport status, and weather information through these resources. All of these information streams provide better Situational Awareness (SA) and a better understanding of the relationship of UAS to other aircraft movements for remote pilots, which enables more efficient operations. Furthermore, other enabling technologies are presented in this paper, such as vision technologies, control techniques, and specific pilot alerts. Finally, the authors have proposed an approach that would include additional information to remote pilots' flight control cockpit-like displays.

% paper number 21
% Considered Large Aircraft - X
% Considered Evaluation of ATCo Workload - X
% Considered Different levels of Familiarity - X
% Considered Mixed Aircraft (UAS and Traditional Aircraft) - √
\par In \cite{wang2017integration}, the authors present the advantages and disadvantages of four architecture alternatives for enabling FAA Next-Gen National Voice System (NVS), which are Legacy Analog, UAS Gateway Connected to Remote Radio Node (RRN), UAS Gateway Connected to AVN and UAS Gateway over FAA Telecommunication Infrastructure (FTI). Considering the architecture choice, UAS Gateway design and functional requirements development are presented. As UAS technology advances and operations become feasible and cost-effective, architectures that support seamless interaction between UAS and the ATC are needed. These architectures should include a UAS network Gateway for managing Air Traffic Voice Node (AVN) within the airspace via a networked Ground-to-Ground (G/G) connection. Several functional requirements must be considered in this context, such as latency, security, access, communication, frequency, and fault. On the other hand, the main components of the NVS include the ATC Voice Node (AVN), which connects the pilot and ATC, and Local Radios (LRs), which are used in tower operations. Finally, as current technologies adopted in UAS operations introduce long latency and may sometimes be unavailable, enabling UAS integration into the NextGen voice system is important. In conclusion, the authors highlight that the 1-to-1 deployment of UAS Gateways to AVN and the deployment of “access gateways” to provide a point of entry for the UAS PIC is the recommended option. However, although this research is an important contribution in terms of integration of UAS considering appropriate communication, the relationship of these aircraft with the ATC is not considered.

\par Finally, this section presented the works related to UAS integration in the National Airspace System (NAS). Each research covers different aspects, and Table \ref{tab:relatedWorksUAS integration} summarizes characteristics of all works based on the following classifications:
\begin{itemize}
    \item \textbf{Large Aircraft (LA):} Indicates if the research considered large UAS in the proposed approach;
    \item \textbf{Impacts on ATCo Workload (ATCoW):} Indicates if the impacts related to UAS operation on ATCo workload are considered;
    \item \textbf{Levels of Familiarity (LF):} Indicates if the proposed integration approach takes the familiarity of ATCo with the particular aircraft into account;
    \item \textbf{Mixed Aircraft (MixA):} Indicates if UAS operations are considered along with Manned Aircraft (MA) operations.
\end{itemize}

\par This table shows that most related works consider a mix of manned and unmanned aircraft. Furthermore, many works consider UAS as a large aircraft. On the other hand, although the impacts of UAS on ATC performance are important to be measured and reduced, only two related works consider the integration from the ATCo perspective. Also, none of the listed works treat all the criteria presented in the Table.

\begin{table}[H]
\centering
\caption{Review of UAS Integration in the National Airspace System (NAS).}
\label{tab:relatedWorksUAS integration}
\begin{tabular}{c|c|
>{\columncolor[HTML]{FD6864}}c |
>{\columncolor[HTML]{FD6864}}c |
>{\columncolor[HTML]{009901}}c }
\hline\hline
% \cellcolor[HTML]{9B9B9B}{\color[HTML]{FFFFFF} Related Work} & \cellcolor[HTML]{9B9B9B}{\color[HTML]{FFFFFF} LA} & \cellcolor[HTML]{9B9B9B}{\color[HTML]{FFFFFF} ATCoW} & \cellcolor[HTML]{9B9B9B}{\color[HTML]{FFFFFF} LF} & \cellcolor[HTML]{9B9B9B}{\color[HTML]{FFFFFF} MixA} \\ \hline
{Related Work} & { LA} & \cellcolor[HTML]{FFFFFF}{ATCoW} & \cellcolor[HTML]{FFFFFF}{LF} & \cellcolor[HTML]{FFFFFF}{MixA} \\ \hline\hline
\cite{Shmelova2016modelling}                                            & \cellcolor[HTML]{FD6864}{\color[HTML]{FFFFFF} X}  & {\color[HTML]{FFFFFF} X}                             & {\color[HTML]{FFFFFF} X}                          & {\color[HTML]{FFFFFF} \checkmark}                            \\ \hline
\cite{pastor2014real}                                              & \cellcolor[HTML]{FD6864}{\color[HTML]{FFFFFF} X}  & \cellcolor[HTML]{009901}{\color[HTML]{FFFFFF} \checkmark}     & \cellcolor[HTML]{009901}{\color[HTML]{FFFFFF} \checkmark}  & \cellcolor[HTML]{FD6864}{\color[HTML]{FFFFFF} X}    \\ \hline
\cite{allignol2016integration}                                             & \cellcolor[HTML]{FD6864}{\color[HTML]{FFFFFF} X}  & {\color[HTML]{FFFFFF} X}                             & {\color[HTML]{FFFFFF} X}                          & {\color[HTML]{FFFFFF} \checkmark}                            \\ \hline
\cite{korn2012stepwise}                                                 & \cellcolor[HTML]{009901}{\color[HTML]{FFFFFF} \checkmark}  & {\color[HTML]{FFFFFF} X}                             & {\color[HTML]{FFFFFF} X}                          & {\color[HTML]{FFFFFF} \checkmark}                            \\ \hline
\cite{de2016approach}                                               & \cellcolor[HTML]{009901}{\color[HTML]{FFFFFF} \checkmark}  & {\color[HTML]{FFFFFF} X}                             & {\color[HTML]{FFFFFF} X}                          & {\color[HTML]{FFFFFF} \checkmark}                            \\ \hline
\cite{branch2016design}                                               & \cellcolor[HTML]{FD6864}{\color[HTML]{FFFFFF} X}  & {\color[HTML]{FFFFFF} X}                             & {\color[HTML]{FFFFFF} X}                          & {\color[HTML]{FFFFFF} \checkmark}                            \\ \hline
\cite{ricardo2014guidelines}                                             & \cellcolor[HTML]{009901}{\color[HTML]{FFFFFF} \checkmark}  & {\color[HTML]{FFFFFF} X}                             & {\color[HTML]{FFFFFF} X}                          & {\color[HTML]{FFFFFF} \checkmark}                            \\ \hline
\cite{ramalingam2011integration}                                          & \cellcolor[HTML]{009901}{\color[HTML]{FFFFFF} \checkmark}  & {\color[HTML]{FFFFFF} X}                             & {\color[HTML]{FFFFFF} X}                          & {\color[HTML]{FFFFFF} \checkmark}                            \\ \hline
\cite{kamienski2015atc}                                        & \cellcolor[HTML]{009901}{\color[HTML]{FFFFFF} \checkmark}  & {\color[HTML]{FFFFFF} X}                             & {\color[HTML]{FFFFFF} X}                          & {\color[HTML]{FFFFFF} \checkmark}                            \\ \hline
\cite{mcfadyen2016terminal}                                            & \cellcolor[HTML]{FD6864}{\color[HTML]{FFFFFF} X}  & {\color[HTML]{FFFFFF} X}                             & {\color[HTML]{FFFFFF} X}                          & {\color[HTML]{FFFFFF} \checkmark}                            \\ \hline
\cite{clothier2017making}                                            & \cellcolor[HTML]{FD6864}{\color[HTML]{FFFFFF} X}  & {\color[HTML]{FFFFFF} X}                             & {\color[HTML]{FFFFFF} X}                          & {\color[HTML]{FFFFFF} \checkmark}                            \\ \hline
\cite{washington2017managing}                                          & \cellcolor[HTML]{009901}{\color[HTML]{FFFFFF} \checkmark}  & {\color[HTML]{FFFFFF} X}                             & {\color[HTML]{FFFFFF} X}                          & {\color[HTML]{FFFFFF} \checkmark}                            \\ \hline
\cite{romero2017proposal}                                              & \cellcolor[HTML]{FD6864}{\color[HTML]{FFFFFF} X}  & {\color[HTML]{FFFFFF} X}                             & {\color[HTML]{FFFFFF} X}                          & {\color[HTML]{FFFFFF} \checkmark}                            \\ \hline
\cite{grimaccia2017risk}                                        & \cellcolor[HTML]{FD6864}{\color[HTML]{FFFFFF} X}  & \cellcolor[HTML]{009901}{\color[HTML]{FFFFFF} \checkmark}     & {\color[HTML]{FFFFFF} X}                          & {\color[HTML]{FFFFFF} \checkmark}                            \\ \hline
\cite{perrottet2017enabling}                                                  & \cellcolor[HTML]{009901}{\color[HTML]{FFFFFF} \checkmark}  & {\color[HTML]{FFFFFF} X}                             & {\color[HTML]{FFFFFF} X}                          & {\color[HTML]{FFFFFF} \checkmark}                            \\ \hline
\cite{sesso2016approach}                                            & \cellcolor[HTML]{009901}{\color[HTML]{FFFFFF} \checkmark}  & {\color[HTML]{FFFFFF} X}                             & {\color[HTML]{FFFFFF} X}                          & {\color[HTML]{FFFFFF} \checkmark}                            \\ \hline
\cite{oztekin2012development}                                             & \cellcolor[HTML]{009901}{\color[HTML]{FFFFFF} \checkmark}  & {\color[HTML]{FFFFFF} X}                             & {\color[HTML]{FFFFFF} X}                          & {\color[HTML]{FFFFFF} \checkmark}                            \\ \hline
\cite{heisey2013reference}                                              & \cellcolor[HTML]{009901}{\color[HTML]{FFFFFF} \checkmark}  & {\color[HTML]{FFFFFF} X}                             & {\color[HTML]{FFFFFF} X}                          & {\color[HTML]{FFFFFF} \checkmark}                            \\ \hline
\cite{wargo2017enhancing}                                            & \cellcolor[HTML]{FD6864}{\color[HTML]{FFFFFF} X}  & {\color[HTML]{FFFFFF} X}                             & {\color[HTML]{FFFFFF} X}                          & {\color[HTML]{FFFFFF} \checkmark}                            \\ \hline
\cite{wang2017integration}                                                & \cellcolor[HTML]{FD6864}{\color[HTML]{FFFFFF} X}  & {\color[HTML]{FFFFFF} X}                             & {\color[HTML]{FFFFFF} X}                          & {\color[HTML]{FFFFFF} \checkmark}                            \\ \hline
\end{tabular}
\end{table}

\section{Simulation of UAS in the Airspace}
\label{sec:simulation}
\par This section presents works related to airspace simulation methods that may include UAS. To identify research gaps, many aspects are analyzed. The works presented in this section are selected as related works according to the presence of the following aspects: the presence of UAS (UAS), Cognitive Impact of Different Aircraft (CIDA), Bad Weather Conditions (BWC), Conflicts Avoidance (CA), Air Traffic Controller (ATCo), and vectoring (Vc) and workload (Wl) evaluation.

% paper number 1
% Considered UAS - √
% Considered TML - x
% Considered Cumulonimbus - x
% Considered Conflicts avoidance - √
% Considered ATCo - x
% Considered Vectoring - x
% Considered Workload - x
\par In \cite{mcfadyen2016simulation}, the authors present two simulation tools focused on unmanned aircraft operations within shared airspace, considering the safety perspective. To accomplish this, a fast pair-wise encounter generator is proposed to simulate the See and Avoid environment, which is demonstrated through statistical performance evaluation of an autonomous See and Avoid decision and control strategy collected in experiments. Also, an unmanned aircraft mission generator is presented, which helps to visualize the impact of multiple unmanned operations. The authors intend to use these analysis tools in exploring some of the fundamental and challenging problems faced by civilian unmanned aircraft system integration and consider that these simple simulation tools can be valuable when assessing a future aerospace environment. Finally, future works, such as applying their strategy in random walk style missions, are pointed out. However, this work does not include Air Traffic Controller (ATCo) aspects in simulation, such as workload. Also, autonomous aircraft do not present a relative cost due to the lack of familiarity with the airspace operators (e.g., ATCo) present with this new technology.

% paper number 2
% Considered UAS - x
% Considered TML - x
% Considered Cumulonimbus - x
% Considered Conflicts avoidance - √
% Considered ATCo - √
% Considered Vectoring - x
% Considered Workload - x
\par Scala et al. \cite{scala2016optimization} propose a methodology for developing an airport arrival and departure manager tool. Optimization and simulation techniques are employed for improving the robustness of the solution. The main goal is to help air traffic controllers manage the inbound and outbound traffic without incurring conflicts or delays, i.e., this tool can help them make the right decisions quickly. The decisions taken in the present methodology for each aircraft are related to entry time and entry speed in the airspace and pushback time at the gate. Finally, this approach presents a smooth flow of aircraft both in the airspace and on the ground. The experiments, which considered the Paris Charles de Gaulle Airport as the case study, showed that conflicts were sensibly reduced. However, although the number of conflicts is reduced in this simulation tool considering this approach, Unmanned Aircraft Systems (UASs) are not considered. Also, the uncertainty related to autonomous control systems (such as the airport arrival and departure manager tool) is not considered.

% paper number 3
% Considered UAS - x
% Considered TML - x
% Considered Cumulonimbus - x
% Considered Conflicts avoidance - √
% Considered ATCo - √
% Considered Vectoring -x 
% Considered Workload - √
\par Farlik \cite{farlik2015conceptual} proposes the concept of air force simulator operational architecture. Considering that live military training in the airspace is expensive and that information technologies have evolved in the past years, simulation becomes a feasible alternative in training building military simulation centers with a high level of realism may be useful in this sense. To train a wide spectrum of personnel together (e.g., pilots and ATCos), simulation capabilities are merged into a single robust simulation environment, which considers, for instance, cooperation. Finally, although the simulation of air defense operations with all its aspects is a complex process, this paper stated the essential conceptual operational architecture of the proposed air defense simulator, helping to structure future simulator architecture according to military requirements. However, this paper presents a set of recommendations but does not include the UAS presence, which may be feasible in the future.

% paper number 4
% Considered UAS - x
% Considered TML - x
% Considered Cumulonimbus - x
% Considered Conflicts avoidance - √
% Considered ATCo - √
% Considered Vectoring - √
% Considered Workload - √
\par The authors in \cite{hu2016simulation} present a simulation study on Air Traffic Control (ATC) strategies aiming to use global traffic information to improve the decision-making process in local ATC. Considering that ATC is a decentralized system, the control sectors face the challenge of using all available information to manage local air traffic safely, smoothly, and cost-effectively. The strategy adopted means how to define and apply various local ATC rules (e.g., first-come-first-served rule) to the coming-in traffic, i.e., traffic that will enter the local sector and whose information is available in global ATC. Finally, a simple ATC network model is set up, and a software simulation system is proposed the simulation results showed that applying an inappropriate set of rules can cause heavy traffic congestion, while an appropriate strategy (based on global information) can significantly reduce delays, improve safety, and increase the efficiency of using the airspace. The authors also indicate future directions of this research, such as introducing more ATC rules and studying the effect of each of them, collaborating with the ATC industry to modify and improve the simulation systems, and designing proper ATC strategies. However, although the proposed strategy considers the ATC, UAS operation is not considered. Also, different costs, in terms of workload, for different aircraft are not considered.

% paper number 5
% Considered UAS - x
% Considered TML - x
% Considered Cumulonimbus - √
% Considered Conflicts avoidance - √
% Considered ATCo - √
% Considered Vectoring - x
% Considered Workload - x
In \cite{mehlitz2016race}, the authors present a framework for facilitating rapid development and deployment of distributed simulations based on Java virtual machines called Runtime for Airspace Concept Evaluation (RACE). Developing large, distributed, human-in-the-loop airspace simulations maybe not be simple, including sophisticated interactive visualization. This framework utilizes the actor programming model and open-source components (e.g., Akka and WorldWind). Finally, the authors highlight three main contributions, which are the provision by actors of the basis for extensibility and scalability, the seamless combination of functional programming (scala) and actors, and the minimal core of RACE with the maturity of its third-party system basis allow applications that go beyond simple simulations. However, although this framework allows adaptations and extensions, it does not consider the UAS presence and its interaction with Air Traffic Controllers (ATCo). Finally, this research is not focused on the final sector of Terminal Manoeuvring Areas (TMA).

% paper number 6
% Considered UAS - x
% Considered TML - x
% Considered Cumulonimbus - x
% Considered Conflicts avoidance - x
% Considered ATCo - √
% Considered Vectoring - x
% Considered Workload - √
\par AgentFly system is presented in \cite{volf2015wide} as fast-time simulation tool. The authors state that this tool is appropriate for National Airspace System (NAS)-wide experiments, and the cognitive behavior of Air Traffic Controllers (ATCo) is considered. The United States NAS, for instance, is very complex, and lots of information need to be shared from specific regions to the whole NAS. Also, increases in air traffic lead to impacts on the difficulty faced by ATCo in avoiding unsafe and inefficient operations. An alternative to the real operation is real-time Human-In-The-Loop (HITL) simulation, which provides valuable feedback on the behavior of human operators but presents limited flexibility and high cost. Thus, AgentFly is proposed as a fast-simulation tool and can be used to perform different experiments, varying the number of aircraft, to facilitate the analyses of specific situations in airspace (e.g., conflict avoidance). Several simulations were conducted, and important metrics were measured. The present results showed that this tool is appropriate to be used as a tool for large-scale experiments, providing detailed data for further analysis. However, although this paper considers the Remotely Piloted Aircraft Systems (RPAS) integration as a possible implementation (including many aspects of their operation as landing) and a human behavior model, which considers workload, additional cognitive workload related to UAS is not taken into account. Finally, the workload during arrival segment execution is not computed using vectoring points and different costs in terms of workload if considering different aircraft (e.g., traditional aircraft and UAS).

% paper number 7
% Considered UAS - x
% Considered TML - x
% Considered Cumulonimbus - √
% Considered Conflicts avoidance - √
% Considered ATCo - √
% Considered Vectoring - √
% Considered Workload - x
\par The authors in \cite{hoekstra2016bluesky} aim to make ATM research results more comparable by sharing tools and data using a fully open-source and open-data approach to air traffic simulation. The main challenges were achieving high fidelity (e.g., aircraft performance) and increasing the community's adoption by keeping the program as simple as possible. Considering the adoption of this platform by many users, this can be considered a useful tool in innovation and application development (e.g., providing specific methods for different problems). The paper describes the difficulties faced when using a fully open-data and open-source policy in this area. However, this work does not consider the UAS presence and its impacts on the ATCo workload.

% paper number 8
% Considered UAS - x
% Considered TML - x
% Considered Cumulonimbus - x
% Considered Conflicts avoidance - √
% Considered ATCo - x
% Considered Vectoring - √
% Considered Workload - x
\par Tra et al. \cite{martijn2017tramodeling} present conflict rate models to determine the intrinsic safety of airspace designs, which consider conflicts between aircraft in different flight phases. Fast-time simulations were performed for different layered airspace concepts, considering unstructured airspaces. The results indicate that the models can estimate the conflict rate for high traffic densities. When comparing the different layered airspace concepts tested, the model predicted, and the simulation results, a clear safety improvement can be noted when the heading range is reduced. Thus the models can be used to study the effect of airspace design parameters on the safety of airspace concepts. However, although this research considers structured and unstructured airspaces, the presence of UAS is not considered.

% paper number 9
% Considered UAS - x
% Considered TML - x
% Considered Cumulonimbus - √
% Considered Conflicts avoidance - √
% Considered ATCo - x
% Considered Vectoring - x
% Considered Workload - x
\par In \cite{alam2008atoms}, the authors introduce the Air Traffic Operations and Management Simulator (ATOMS), an air traffic and airspace modeling and simulation system for free-flight concepts. This tool simulates end-to-end airspace operations and air navigation procedures for conventional air traffic. A multiagent-based modeling paradigm for modular design and easy integration of various air traffic subsystems is adopted. Also, advanced Air Traffic Management (ATM) concepts that are envisioned in free flight are prototyped in this research, including Airborne Separation Assurance (ASA), Cockpit Display of Traffic Information (CDTI), and weather avoidance. The results showed that advanced ATM concepts present an appropriate scenario for free flights. However, this research does not consider the ATCo workload and the impact of the UAS presence on it. Also, it does not consider a vectoring point-based workload evaluation.

% paper number 10
% Considered UAS - x
% Considered TML - x
% Considered Cumulonimbus - x
% Considered Conflicts avoidance - √
% Considered ATCo - √
% Considered Vectoring - √
% Considered Workload - x
The authors in \cite{pavlinovic2017air} focus on simulation-based Air Traffic Controller (ATCo) training using the Beginning to End for Simulation and Training (BEST), which is a simulation tool adopted in training organizations in Europe. Although the BEST simulator covers all levels and types of training (e.g., basic, validation and conversation refresher), this research is focused on the basic part of the initial training. Furthermore, insights into the challenges the candidates face when mastering the techniques of performance-based training are presented. ATCos are responsible for guiding aircraft through the airspace, and throughout the whole education and later work, their extensive training (which considers practical exercises performed on computer devices) is divided into three phases. They are the initial training (basic and rating training), unit training (transitional, pre-on-the-job, and on-the-job training), and continuation training (conversion and refresher training). Moreover, BEST simulator meets all the objectives and requirements prescribed for basic ATCo training. However, this research does not consider the aspects related to UAS integration (e.g., increase in cognitive workload), and different levels of aircraft in terms of ATCo familiarity are not considered.

% paper number 11
% Considered UAS - x
% Considered TML - x
% Considered Cumulonimbus - √
% Considered Conflicts avoidance - √
% Considered ATCo - x
% Considered Vectoring - x
% Considered Workload - x
Young et al. \cite{young2013modeling} proposes an approach to describe the process of creating and using weather polygons for simulation and analysis activities using the Federal Aviation Administration (FAA)'s Concept Analysis Branch, including an example study focused on weather impacts on flights efficiency and tested in a fast-time simulation environment. Considering that weather has substantial impacts on the National Airspace System (NAS) and that most simulation and analysis tools are unable to represent weather activity effectively, the development of capabilities that benefits some operational improvements cannot be quantified, i.e., the weather may impact negatively on solutions that improve the airspace efficiency consider good weather conditions. The FAA’s Concept Analysis Branch (ANG-C41) developed a tool to create weather polygons, which is a concise model to store and process and can be modeled as restricted airspace that moves across the NAS that considers high-fidelity weather data. This enables the measurement of the impact of operational improvements on weather-related flight delays, and, thus, an analysis of the efficiency of current weather avoidance operations was conducted using weather polygons in algorithms to calculate the distance of each flight from the weather at different severity levels as well as to identify flights which rerouted to avoid moderate to severe convective weather. Finally, capability enables the FAA to represent the impact of convective weather on NextGen operational improvements. However, although this research is broad in terms of weather simulations, this does not include UAS operation in bad weather avoidance. Furthermore, the ATCo workload related to this process is not considered.

% paper number 12
% Considered UAS - x
% Considered TML - x
% Considered Cumulonimbus - √
% Considered Conflicts avoidance - x
% Considered ATCo - √
% Considered Vectoring - x
% Considered Workload - x
\par McGovern et al. \cite{mcgovern2009stochastic} propose and present an overview of a Monte Carlo-based stochastic airspace simulation tool and its formulation as programming languages, environments, and other development tools. The main objective is to provide a documented, lifecycle-managed, multi-processor capable, stochastic simulation capability to enable the analysis of procedures and equipment for aircraft flight into shared airspace. Thus, the selection, design, and implementation of the mathematical models and verification and validation processes are conducted. Since real experiments are expensive and unfeasible, modeling and simulation are often used to study the physical world. Furthermore, navigation aids, surveillance systems, pilots, aircraft, Air Traffic Controllers (ATCos), and weather are desirable features for a useful simulation tool, and in this research, the authors consider all of them and a Graphical User Interface (GUI) integrating world-wide photo-realistic airport depictions and real-time three-dimensional animation. Finally, this paper focuses on the interaction of components in shared airspace, and the software tool and its formulation are presented. However, this work does not consider the UAS integration into airspace and its impacts on ATCo operation as well as the ATCo workload.

% paper number 13
% Considered UAS - √
% Considered TML - x
% Considered Cumulonimbus - √
% Considered Conflicts avoidance - √
% Considered ATCo - √
% Considered Vectoring - √
% Considered Workload - x
\par The authors in \cite{homola2016uas} developed a simulation component, considering the UAS Traffic Management (UTM) paradigm, that supports near and long-term live flight testing and exploration. The capabilities of the simulation tool are depicted in this work. In this context, NASA has started to work collaboratively in research with the Federal Aviation Administration (FAA) and other stakeholders (government, industry, and academia) to explore the concepts and requirements of safe and scalable operations of small Unmanned Aircraft Systems (UAS) in low-altitude airspaces. Finally, a powerful research and development platform capable of addressing the multitude of questions is developed considering that the UTM laboratory is ideally suited to progress the state of UTM research and knowledge. However, although this research considers the UAS presence and its control, large aircraft (e.g., traditional aircraft) are not considered. Also, the control of manned and unmanned aircraft is performed by different agents, and the aircraft are not included in shared high-altitude airspace.

% paper number 14
% Considered UAS - x
% Considered TML - x
% Considered Cumulonimbus - x
% Considered Conflicts avoidance - √
% Considered ATCo - √
% Considered Vectoring - x
% Considered Workload - x
\par Different utilization modes of Closely Spaced Parallel Runways (CSPRs), which are employed in the construction of parallel runways, are analyzed in \cite{li2015utilization}, considering different thresholds. This analysis is conducted using the simulation software SIMMOD, which was applied to build simulation models for different utilization modes of runways with a staggered threshold. This systematic analysis aimed to evaluate airport capacity and operational efficiency quantitatively. The authors showed through experiments that, considering the existing air traffic control operation rules, CSPRs usage enables the airspace to support from 765 to 815 movements on each peak day. Also, 55 movements are supported in each peak hour. If the runway threshold is staggered in terms of the approach direction and a bypass taxiway is provided, i.e., considering adaptations of the runway due to air traffic state (e.g., dynamic change to reduce arrival delay), the mode landing on the inside runway and taking-off from the outside runway shows up as the most efficient mode, which increases the operation efficiency by about 5\%. However, this research explores the capabilities of a widely used simulation tool to build simulation models. Also, the focus is on evaluating landings and take-offs, which is different from the final approach, although it substantially impacts this previous phase. Finally, the presence of autonomous control systems and their impacts on personnel performance are not considered.

% paper number 15
% Considered UAS - √
% Considered TML - x
% Considered Cumulonimbus - x
% Considered Conflicts avoidance - √
% Considered ATCo - x
% Considered Vectoring - x
% Considered Workload - x
\par In \cite{sun2009new}, a new 4D trajectory generation method is proposed. This method is based on historical radar data processing, considering traffic flow theory to generate the flight states and introducing the multiple interacting models smoother and spline interpolation to determine the intermediate flying status. 4D trajectory generation, one of the most fundamental functions in the airspace simulation system, is currently based on the partitioning of the flight, i.e., the entire flight is divided into several parts, enabling the usage of models to generate the flying states. However, in the method proposed in this research, the problems of generating the initial state of an aircraft and depicting the 4D flying status are addressed. The results presented in this work, obtained from the simulated trajectories and real trajectories by the MATLAB software, showed that the method is valid and practical. However, the authors do not consider the interaction of these aircraft with the ATCo. Also, problems related to workload aspects, as well as bad weather conditions, are not faced. Finally, although the authors consider real data, the standard curve rate employed in aviation is not highlighted.

% paper number 16
% Considered UAS - √
% Considered TML - x
% Considered Cumulonimbus - x
% Considered Conflicts avoidance - x
% Considered ATCo - x
% Considered Vectoring - x
% Considered Workload - x
\par Bucceroni et al. \cite{bucceroni2016multi} propose a system for integrating Unmanned Aerial System (UAS) visual observers into the distributed simulation capability of the Federal Aviation Administration (FAA). This distributed simulation, which employs large-scale virtual reality systems, is used to demonstrate terrain surrounding flight tests in virtual environments and generate the observer’s views (ground- and air-based). Three situations are considered: stationary ground-based monitoring, mobile air-based monitoring, and seaborne monitoring. As large-scale distributed visualizations are routinely used by organizations (industry and government), this approach is beneficial and has considerable importance in the research associated with the FAA’s current work on adapting itself to Next Generation Air Transportation System, considering that this new system will include UAS into shared airspace. Thus, locally caching real-world terrain with data provided was integrated into a graphical interface to give the operator the UAS position and information from multiple perspectives in a distributed simulation. However, the simulation tool proposed in this research does not include interaction between the aircraft operator and the Air Traffic Controller (ATCo). Finally, this is a work conducted aiming at the future integration of UAS into shared airspace, but it does not consider the impacts these autonomous systems have on personnel performance (e.g., workload).

% paper number 17
% Considered UAS - √
% Considered TML - x
% Considered Cumulonimbus - √
% Considered Conflicts avoidance - √
% Considered ATCo - x
% Considered Vectoring - x
% Considered Workload - x
\par Borener et al. \cite{borener2015modeling} present Unmanned Aircraft Systems (UAS) modeling and simulation, which consider a use case scenario that is consistent with the FAA’s concept of operations for integration of UAS into shared airspace and employs sensing (using actual radar track traffic data) and medium fixed-wing UAS. The proposed simulations offer functionality related to UAS operations, such as ‘detect and avoid’, mission profiles, positional variance, performance variance, fuzzy conflicts, variation in time spent in communication, and deviation from planned or intent profiles. Based on the RAMS plus fast-time simulator tool, the simulations aim to evaluate the separation indices and the number and severity of the separation events. The experiments, conducted in a simulated Houston Metroplex Environment, showed that multiple UAS would considerably increase the likelihood of separation events and separation critically indices and the usage of the "return to departure land site" contingency operation in case of failures in UAS communication link has a considerable impact on separation events. A difficulty faced by researchers, though, is the lack of historical data on UAS operations. However, this research does not include the interaction between UAS and ATCo; consequently, the workload is not considered.

% paper number 18
% Considered UAS - x
% Considered TML - x
% Considered Cumulonimbus - √
% Considered Conflicts avoidance - √
% Considered ATCo - √
% Considered Vectoring - √
% Considered Workload - √
The authors in \cite{kopardekar2009airspace} describe the results of a Dynamic Density (DD) human-in-the-loop simulation. The DD model adopted aims to measure the complexity of a given airspace area, and measures presented at the US/Europe ATM 2003 Seminar were used in this research. Thus, the simulation included Reduced Vertical Separation on the Cleveland Air Route Traffic Control Centre’s airspace, and the considered traffic was actively controlled throughout the simulation. Due to the difficulty related to real-world simulation, i.e., it may not be feasible to conduct experiments with real aircraft, simulations were adopted in this research. One should note that human-in-the-loop simulations may offer more accurate data on, for instance, airspace capacity evaluation. The simulated experiments employed six Certified Professional Controllers (CPCs) and one Operations Supervisor from Cleveland and were conducted in the high-fidelity Display System Replacement (DSR) Laboratory. The experiments showed the DD metric performed better than the aircraft count, which is a usual complexity measure. However, although this research has valuable results in simulation and airspace capacity and is considered specialists in Air Traffic Control (ATC), UAS are not considered in the approach. Finally, cognitive factors (e.g., lack of familiarity with ATCo with a specific aircraft) are not considered.

% paper number 19
% Considered UAS - √
% Considered TML - x
% Considered Cumulonimbus - √
% Considered Conflicts avoidance - √
% Considered ATCo - √
% Considered Vectoring - x
% Considered Workload - √
In \cite{horio2012safety}, the LMI ACAS (Airspace Conflict Analysis Simulation) tool is presented. This 3-dimensional simulation modeling tool, and its application meets the analytical requirements. The benefits of implementing a multidimensional visualization are also presented. The conducted case study, which employs the ACAS tool for a safety risk assessment based on conflict probability, demonstrates the capacity of the framework to evaluate safety risk. Also, a set of concerns that include traffic growth, Next Generation Air Transportation System (NextGen) technologies, dynamic airspace sector reconfiguration, and the integration of UAS into shared airspace are considered. As NextGen is under development, modeling complex, diverse future scenarios is important, thus, better solutions can be provided in terms of specific metrics (e.g., safety). Some causes and consequences related to risk analysis are pointed out. For instance, an increase in passenger demand leads to a higher traffic density. Considering that ACAS is not meant to be a NAS-wide simulation of all aspects of flight, this research proposes an agile tool for exploring NextGen aviation concepts and technologies from the safety perspective. However, this research does not include the cognitive workload associated with special aircraft operations (e.g., UAS).

% paper number 20
% Considered UAS - x
% Considered TML - x
% Considered Cumulonimbus - x
% Considered Conflicts avoidance - √
% Considered ATCo - √
% Considered Vectoring - √
% Considered Workload - x
\par In \cite{hong2016conflict}, numerical simulations are used in order to demonstrate the effectiveness of the proposed conflict management approach, which ensures conflict avoidance among aircraft and the transition of aircraft into adjacent airspace. To avoid conflicts, complexity is modeled as aircraft heading and speed deviations in a given sector. Considering more than one sector, a specific architecture is proposed for planning to minimize complexity for the neighboring sectors. More specifically, the conflict avoidance problem can be seen as a mixed integer Linear Programming (LP) subject to maneuver constraints. Thus, the aircraft can find the optimal solution by solving the LP problem, resolving conflicts among the aircraft, and reducing the air traffic complexity of the neighboring sectors. Moreover, the proposed conflict management algorithm can identify aircraft's optimal conflict resolution maneuver in near real-time considering multi-sector environments. The authors intend to investigate the relationship between maneuver constraints and traffic complexity in future works. However, although this research is interesting from the conflict avoidance perspective, it does not deal with fully autonomous or remotely piloted aircraft. Finally, ATCo operation and the impacts of the proposed approach on his/her workload are not considered.

% Summing up
\par In this section, the works related to airspace simulation were presented. Each work covers different aspects, but to identify the similarities and differences, Table \ref{tab:relatedWorksAirspaceSimulation} presents all works, which are classified as follows:
\begin{itemize}
    \item \textbf{Unmanned Aircraft System (UAS):} Indicates if UAS are considered in a given research. One should note that Remotely Piloted Aircraft Systems (RPAS), as a sort of UAS, are also considered.
    \item \textbf{Cognitive Impact of Different Aircraft (CIDA):} Indicates if the impacts related to special aircraft (e.g., UAS) operation on personnel performance.
    \item \textbf{Bad Weather Conditions (BWC):} Indicates if the proposed simulation tool deals with the challenges imposed by bad weather conditions.
    \item \textbf{Conflicts Avoidance (CA):} Indicates if conflict avoidance is prioritized in the given simulation tool.
    \item \textbf{Air Traffic Controller (ATCo):} Indicates if ATCo operation is one simulation focus.
    \item \textbf{Vectoring (Vc):} Indicates if aircraft are controlled by Vectoring Points (VP).
    \item \textbf{Workload (Wl):} Indicates if the workload of ATCo is evaluated in the simulation tool.
\end{itemize}

\par This table shows that most related works deal with collision avoidance problems in simulation, and many consider the ATCo operation. Few works include weather conditions, vectoring, and workload evaluation in experiments. Thus, UAS appears in a few works. Finally, none of the listed related works treat the cognitive impact on, for example, workload due to the lack of familiarity of ATCo with a new aircraft (e.g., UAS). Also, none of the listed works treats all the criteria presented in the Table.

\begin{table}[H]
\centering
\caption{Review of UAS simulation in the National Airspace System (NAS).}
\label{tab:relatedWorksAirspaceSimulation}
\begin{tabular}{c|
>{\columncolor[HTML]{FD6864}}c |
>{\columncolor[HTML]{FD6864}}c |
>{\columncolor[HTML]{FD6864}}c |
>{\columncolor[HTML]{009901}}c |
>{\columncolor[HTML]{009901}}c |
>{\columncolor[HTML]{FD6864}}c |
>{\columncolor[HTML]{FD6864}}c }
\hline\hline
% \cellcolor[HTML]{9B9B9B}{\color[HTML]{FFFFFF} Related Work} & \cellcolor[HTML]{9B9B9B}{\color[HTML]{FFFFFF} UAS} & \cellcolor[HTML]{9B9B9B}{\color[HTML]{FFFFFF} CIDA} & \cellcolor[HTML]{9B9B9B}{\color[HTML]{FFFFFF} BWC} & \cellcolor[HTML]{9B9B9B}{\color[HTML]{FFFFFF} CA} & \cellcolor[HTML]{9B9B9B}{\color[HTML]{FFFFFF} ATCo} & \cellcolor[HTML]{9B9B9B}{\color[HTML]{FFFFFF} Vc} & \cellcolor[HTML]{9B9B9B}{\color[HTML]{FFFFFF} Wl} \\ \hline
\cellcolor[HTML]{FFFFFF}{Related Work} & \cellcolor[HTML]{FFFFFF}{UAS} & \cellcolor[HTML]{FFFFFF}{CIDA} & \cellcolor[HTML]{FFFFFF}{BWC} & \cellcolor[HTML]{FFFFFF}{CA} & \cellcolor[HTML]{FFFFFF}{ATCo} & \cellcolor[HTML]{FFFFFF}{Vc} & \cellcolor[HTML]{FFFFFF}{Wl} \\ \hline\hline
\cite{mcfadyen2016simulation}                                             & \cellcolor[HTML]{009901}{\color[HTML]{FFFFFF} \checkmark}   & {\color[HTML]{FFFFFF} X}                            & {\color[HTML]{FFFFFF} X}                           & {\color[HTML]{FFFFFF} \checkmark}                          & \cellcolor[HTML]{FD6864}{\color[HTML]{FFFFFF} X}    & {\color[HTML]{FFFFFF} X}                          & {\color[HTML]{FFFFFF} X}                          \\ \hline
\cite{scala2016optimization}                                                & {\color[HTML]{FFFFFF} X}                           & {\color[HTML]{FFFFFF} X}                            & {\color[HTML]{FFFFFF} X}                           & {\color[HTML]{FFFFFF} \checkmark}                          & {\color[HTML]{FFFFFF} \checkmark}                            & {\color[HTML]{FFFFFF} X}                          & {\color[HTML]{FFFFFF} X}                          \\ \hline
\cite{farlik2015conceptual}                                               & {\color[HTML]{FFFFFF} X}                           & {\color[HTML]{FFFFFF} X}                            & {\color[HTML]{FFFFFF} X}                           & {\color[HTML]{FFFFFF} \checkmark}                          & {\color[HTML]{FFFFFF} \checkmark}                            & {\color[HTML]{FFFFFF} X}                          & \cellcolor[HTML]{009901}{\color[HTML]{FFFFFF} \checkmark}  \\ \hline
\cite{hu2016simulation}                                                   & {\color[HTML]{FFFFFF} X}                           & {\color[HTML]{FFFFFF} X}                            & {\color[HTML]{FFFFFF} X}                           & {\color[HTML]{FFFFFF} \checkmark}                          & {\color[HTML]{FFFFFF} \checkmark}                            & \cellcolor[HTML]{009901}{\color[HTML]{FFFFFF} \checkmark}  & \cellcolor[HTML]{009901}{\color[HTML]{FFFFFF} \checkmark}  \\ \hline
\cite{mehlitz2016race}                                             & {\color[HTML]{FFFFFF} X}                           & {\color[HTML]{FFFFFF} X}                            & \cellcolor[HTML]{009901}{\color[HTML]{FFFFFF} \checkmark}   & {\color[HTML]{FFFFFF} \checkmark}                          & {\color[HTML]{FFFFFF} \checkmark}                            & {\color[HTML]{FFFFFF} X}                          & {\color[HTML]{FFFFFF} X}                          \\ \hline
\cite{volf2015wide}                                                 & {\color[HTML]{FFFFFF} X}                           & {\color[HTML]{FFFFFF} X}                            & {\color[HTML]{FFFFFF} X}                           & \cellcolor[HTML]{FD6864}{\color[HTML]{FFFFFF} X}  & {\color[HTML]{FFFFFF} \checkmark}                            & {\color[HTML]{FFFFFF} X}                          & \cellcolor[HTML]{009901}{\color[HTML]{FFFFFF} \checkmark}  \\ \hline

\cite{hoekstra2016bluesky}                                             & {\color[HTML]{FFFFFF} X}                           & {\color[HTML]{FFFFFF} X}                            & \cellcolor[HTML]{009901}{\color[HTML]{FFFFFF} \checkmark}   & {\color[HTML]{FFFFFF} \checkmark}                          & {\color[HTML]{FFFFFF} \checkmark}                            & \cellcolor[HTML]{009901}{\color[HTML]{FFFFFF} \checkmark}  & {\color[HTML]{FFFFFF} X}                          \\ \hline
\cite{martijn2017tramodeling}                                                  & {\color[HTML]{FFFFFF} X}                           & {\color[HTML]{FFFFFF} X}                            & {\color[HTML]{FFFFFF} X}                           & {\color[HTML]{FFFFFF} \checkmark}                          & \cellcolor[HTML]{FD6864}{\color[HTML]{FFFFFF} X}    & \cellcolor[HTML]{009901}{\color[HTML]{FFFFFF} \checkmark}  & {\color[HTML]{FFFFFF} X}                          \\ \hline
\cite{alam2008atoms}                                                & {\color[HTML]{FFFFFF} X}                           & {\color[HTML]{FFFFFF} X}                            & \cellcolor[HTML]{009901}{\color[HTML]{FFFFFF} \checkmark}   & {\color[HTML]{FFFFFF} \checkmark}                          & \cellcolor[HTML]{FD6864}{\color[HTML]{FFFFFF} X}    & {\color[HTML]{FFFFFF} X}                          & {\color[HTML]{FFFFFF} X}                          \\ \hline
\cite{pavlinovic2017air}                                     & {\color[HTML]{FFFFFF} X}                           & {\color[HTML]{FFFFFF} X}                            & {\color[HTML]{FFFFFF} X}                           & {\color[HTML]{FFFFFF} \checkmark}                          & {\color[HTML]{FFFFFF} \checkmark}                            & \cellcolor[HTML]{009901}{\color[HTML]{FFFFFF} \checkmark}  & {\color[HTML]{FFFFFF} X}                          \\ \hline
\cite{young2013modeling}                                                & {\color[HTML]{FFFFFF} X}                           & {\color[HTML]{FFFFFF} X}                            & \cellcolor[HTML]{009901}{\color[HTML]{FFFFFF} \checkmark}   & {\color[HTML]{FFFFFF} \checkmark}                          & \cellcolor[HTML]{FD6864}{\color[HTML]{FFFFFF} X}    & {\color[HTML]{FFFFFF} X}                          & {\color[HTML]{FFFFFF} X}                          \\ \hline
\cite{mcgovern2009stochastic}                                             & {\color[HTML]{FFFFFF} X}                           & {\color[HTML]{FFFFFF} X}                            & \cellcolor[HTML]{009901}{\color[HTML]{FFFFFF} \checkmark}   & \cellcolor[HTML]{FD6864}{\color[HTML]{FFFFFF} X}  & {\color[HTML]{FFFFFF} \checkmark}                            & {\color[HTML]{FFFFFF} X}                          & {\color[HTML]{FFFFFF} X}                          \\ \hline
\cite{homola2016uas}                                               & \cellcolor[HTML]{009901}{\color[HTML]{FFFFFF} \checkmark}   & {\color[HTML]{FFFFFF} X}                            & \cellcolor[HTML]{009901}{\color[HTML]{FFFFFF} \checkmark}   & {\color[HTML]{FFFFFF} \checkmark}                          & {\color[HTML]{FFFFFF} \checkmark}                            & \cellcolor[HTML]{009901}{\color[HTML]{FFFFFF} \checkmark}  & {\color[HTML]{FFFFFF} X}                          \\ \hline
\cite{li2015utilization}                                                   & {\color[HTML]{FFFFFF} X}                           & {\color[HTML]{FFFFFF} X}                            & {\color[HTML]{FFFFFF} X}                           & {\color[HTML]{FFFFFF} \checkmark}                          & {\color[HTML]{FFFFFF} \checkmark}                            & {\color[HTML]{FFFFFF} X}                          & {\color[HTML]{FFFFFF} X}                          \\ \hline
\cite{sun2009new}                                                  & \cellcolor[HTML]{009901}{\color[HTML]{FFFFFF} \checkmark}   & {\color[HTML]{FFFFFF} X}                            & {\color[HTML]{FFFFFF} X}                           & {\color[HTML]{FFFFFF} \checkmark}                          & \cellcolor[HTML]{FD6864}{\color[HTML]{FFFFFF} X}    & {\color[HTML]{FFFFFF} X}                          & {\color[HTML]{FFFFFF} X}                          \\ \hline
\cite{bucceroni2016multi}                                            & \cellcolor[HTML]{009901}{\color[HTML]{FFFFFF} \checkmark}   & {\color[HTML]{FFFFFF} X}                            & {\color[HTML]{FFFFFF} X}                           & \cellcolor[HTML]{FD6864}{\color[HTML]{FFFFFF} X}  & \cellcolor[HTML]{FD6864}{\color[HTML]{FFFFFF} X}    & {\color[HTML]{FFFFFF} X}                          & {\color[HTML]{FFFFFF} X}                          \\ \hline
\cite{borener2015modeling}                                              & \cellcolor[HTML]{009901}{\color[HTML]{FFFFFF} \checkmark}   & {\color[HTML]{FFFFFF} X}                            & \cellcolor[HTML]{009901}{\color[HTML]{FFFFFF} \checkmark}   & {\color[HTML]{FFFFFF} \checkmark}                          & \cellcolor[HTML]{FD6864}{\color[HTML]{FFFFFF} X}    & {\color[HTML]{FFFFFF} X}                          & {\color[HTML]{FFFFFF} X}                          \\ \hline
\cite{kopardekar2009airspace}                                           & {\color[HTML]{FFFFFF} X}                           & {\color[HTML]{FFFFFF} X}                            & \cellcolor[HTML]{009901}{\color[HTML]{FFFFFF} \checkmark}   & {\color[HTML]{FFFFFF} \checkmark}                          & {\color[HTML]{FFFFFF} \checkmark}                            & \cellcolor[HTML]{009901}{\color[HTML]{FFFFFF} \checkmark}  & \cellcolor[HTML]{009901}{\color[HTML]{FFFFFF} \checkmark}  \\ \hline
\cite{horio2012safety}                                                & \cellcolor[HTML]{009901}{\color[HTML]{FFFFFF} \checkmark}   & {\color[HTML]{FFFFFF} X}                            & \cellcolor[HTML]{009901}{\color[HTML]{FFFFFF} \checkmark}   & {\color[HTML]{FFFFFF} \checkmark}                          & {\color[HTML]{FFFFFF} \checkmark}                            & {\color[HTML]{FFFFFF} X}                          & \cellcolor[HTML]{009901}{\color[HTML]{FFFFFF} \checkmark}  \\ \hline
\cite{hong2016conflict}                                                 & {\color[HTML]{FFFFFF} X}                           & {\color[HTML]{FFFFFF} X}                            & {\color[HTML]{FFFFFF} X}                           & {\color[HTML]{FFFFFF} \checkmark}                          & {\color[HTML]{FFFFFF} \checkmark}                            & \cellcolor[HTML]{009901}{\color[HTML]{FFFFFF} \checkmark}  & {\color[HTML]{FFFFFF} X}                          \\ \hline
\end{tabular}
\end{table}

\section{Arrival segment optimization considering UAS}
\label{sec:arrival}
\par This section shows a literature review conducted toward optimization approaches for arrival segment design considering the UAS presence. The proposals are analyzed from different perspectives. To select and classify the related works present in this section, the consideration of following aspects of each approach are taken into account: National Airspace System (NAS), Final Arrival Segment Design (FASD), Complex Situations (CS), Bad Weather Conditions (BWC), Minimum Separation (MS), UAS presence (UAS), and Time as a Constraint (TC).

% Paper 1
% non-segregated airspace - √
% Initial phase of final approach - X
% Complex situations - √
% Bad weather conditions - X
% Minimum Separation - √
% UAS presence - X
% Time as a constraint - √
\par Alonso-Ayuso et al. \cite{alonso2016multiobjective} presents an approach that employs a mixed integer linear approximation to a Mixed Integer Nonlinear Optimization (MINO) model for the conflict resolution problem in air traffic management, i.e., for providing aircraft configurations to avoid conflicts, which is the loss of the minimum separation between two given aircraft. The problem is solved by considering an initial position of a set of aircraft and applying changes to their position, velocity, and heading angles. Thus, a multi-criteria scheme and a Sequential Mixed Integer Linear Optimization (SMILO) approach are also presented. This is due to the achievement of solutions in a short computing time. Furthermore, a comparison between the results obtained by using the state-of-the-art MINO solvers and SMILO performance in a broad testbed is also considered, which showed that both presented similar solutions, but the proposed approach requires a very short computing time. Finally, the authors highlight that for large-size instances (e.g., above five aircraft), the computing time is higher than the one required by real-life operational applications and that other meta-heuristics can reduce even the computing time without deteriorating the SMILO solution as a future research line. However, this research does not consider the operation of UAS in the NAS.

% Paper 2
% non-segregated airspace - √
% Initial phase of final approach - X
% Complex situations - √
% Bad weather conditions - X
% Minimum Separation - √
% UAS presence - X
% Time as a constraint - X
\par The authors in \cite{gao2012cooperative} present a cooperative multi-aircraft Conflict Resolution (CR) method based on co-evolution. The paths are composed of sub-populations considered in a Particle Swarm Optimization (PSO) implementation, in which the fitness is evaluated by cooperation among individuals from different sub-population and is adopted for its advantages such as fewer parameters and computation and faster convergence. One should note that each particle is seen as a point of D-dimension space. Further, an encoding method with an adaptive searching mechanism is introduced to improve the searching efficiency. Compared with Genetic Algorithms (GA) currently being used for conflict resolution path optimization, the results achieved by this approach achieved higher system efficiency, which is a manner to measure how similar a given path is to the smallest possible path. Considering 2, 4, and 6 aircraft, the proposed approach outperformed the GA approach. However, although this research employs the PSO successfully, this research does not consider the fitness evaluation and particle update processes to be conducted in parallel for each particle, which can improve the performance considerably. Also, bad weather conditions are not taken into account. Finally, the UAS presence and arrival segments definition are not considered.

% Paper 3
% non-segregated airspace - √
% Final approach optimization - √
% Complex situations - √
% Bad weather conditions - X
% Minimum Separation - √
% UAS presence - X
% Time as a constraint - X
\par Ahmed et al. \cite{ahmed2017evolutionary} present an evolutionary method for optimizing the aircraft path planning algorithm in Terminal Maneuvering Area (TMA). This method, which provides near-optimal aircraft arrival sequences, aims to deliver the aircraft to the Final Approach Fix (FAF). The paths are built to conduct the aircraft from the Initial Approach Fix (IAF) to the FAF considering intermediates waypoints called Intermediate Fix (IF). The classic Genetic Algorithm (GA)-based optimization technique with conflict detection and resolution used in this effort minimizes the inter-arrival time. Furthermore, conflict-free path planning for an Air Traffic Controller (ATC) is also obtained. One should note that conflict between any two aircraft is detected based on their future arrival time at the waypoint. The results show that the proposed approach provides a near-optimal solution compared to the traditional GA-based algorithm, which does not consider airspace constraints (e.g., speed). 

% Paper 4
% non-segregated airspace - √
% Initial phase of final approach - X
% Complex situations - √
% Bad weather conditions - X
% Minimum Separation - √
% UAS presence - X
% Time as a constraint - √
\par In \cite{sama2017scheduling}, the authors proposed a mixed integer linear programming formulation to optimize, in real-time, the take-off and landing operations at a busy Terminal Maneuvering Area (TMA) in case of traffic congestion by investigating the trade-off aspects between performance indicators of practical interest. This method also considers safety constraints with high precision. As TMAs are becoming problematic, especially in the major European airports, since there is a limited possibility of building new infrastructures, alternative solutions (e.g., optimization models) are desired. The real-time problem of effectively managing aircraft operations is challenging, especially due to the inclusion of safety regulations into the optimization model and several performance indicators. This inclusion leads to achieving feasible and reasonable solutions in terms of safety and efficiency, even considering that there is no well-recognized objective function and traffic controllers often use simple scheduling rules. The experiments were performed considering simulated scenarios in the two major Italian airports, Milano Malpensa and Roma Fiumicino. In this context, random landing and take-off aircraft disturbances are built. In the optimization process, practical-size instances are solved to (near) optimality by employing a commercial solver. Finally, a computational analysis enables the selection of solutions that presents considerable quality in balancing the various indicators trade-off. However, this research focuses on scheduling and does not consider the presence of UAS and the impact of the inclusion of this new technology into the ATC. 

% Paper 5
% non-segregated airspace - √
% Initial phase of final approach - √
% Complex situations - √
% Bad weather conditions - X
% Minimum Separation - √
% UAS presence - X
% Time as a constraint - X
\par Sam{\`a} et al. \cite{sama2017optimal} deals with the TMA aircraft scheduling problem, which requires conflict-free schedules for all aircraft, whereas the overall aircraft delays are minimized. Furthermore, this research also deals with the aircraft landing trajectory optimization problem, which requires a landing trajectory that minimizes the travel time or the fuel consumption for each aircraft. In this context, a framework for the lexicographic optimization of both problems is proposed, which solves the two problems sequentially based on a defined lexicographic order of importance for the performance indicators, i.e., the most important performance indicator defines the first problem to be optimized. Note that the second problem is solved considering some outputs of the solution of the first problem. The experiments, performed on simulated Milano Malpensa airport instances and considering different optimization lexicographic orders and performance indicators, show the existence of performance gaps between the optimized indicators of the two problems, highlighting the multi-objective nature of the problem when different lexicographic optimization approaches are considered. However, this research does not consider some aspects, such as bad weather conditions.

% Paper 6
% non-segregated airspace - √
% Initial phase of final approach - √
% Complex situations - √
% Bad weather conditions - X
% Minimum Separation - √
% UAS presence - X
% Time as a constraint - √
\par A number of algorithmic improvements implemented in the AGLIBRARY solver, a state-of-the-art optimization solver for complex routing and scheduling problems, to improve the possibility of finding good quality solutions quickly, is presented in \cite{sama2017metaheuristics}. Intelligent decision support systems for the real-time management of landing and take-off operations, which can be effective in helping ATCos at busy Terminal Control Areas (TMAs), aim to optimize aircraft sequencing. This problem, which can be faced as a mixed integer linear program, is strongly NP-hard, and heuristic algorithms are typically adopted in practice to compute good quality solutions in a short computation time. In this context, the framework proposed in this paper starts from a feasible initial solution for the aircraft scheduling problem with fixed routes, computed via a truncated branch-and-bound algorithm, and, further, metaheuristics (e.g., variable neighborhood search, tabu search, and hybrid schemes) are applied to improve the solution by re-routing some aircraft in the TMA. Finally, the results showed that the metaheuristics quickly achieve solutions of remarkable quality compared with a commercial solver. However, parallel implementations of the metaheuristics, which may reduce the execution time considerably, are not considered. Finally, UAS presence is not taken into account.

% Paper 7
% non-segregated airspace - X
% Initial phase of final approach - X
% Complex situations (many large aircraft) - X
% Bad weather conditions - √
% Minimum Separation - X
% UAS presence - √
% Time as a constraint - √
\par The authors in \cite{silva2017heuristic} apply heuristic and genetic algorithms approach for the path planning problem for UAVs. This approaches consider emergency landing procedures and aim to mitigate the probability of reaching unsafe situations. The path re-planning, caused by several factors such as equipment failures and leads missions to be aborted by executing an emergency planned landing, is introduced through a mathematical formulation. In this context, path planning approaches that employ greedy heuristic, which aims to find feasible paths quickly, genetic algorithm, and multi-population genetic algorithm, which tends to return better quality solutions, are introduced and evaluated considering a large set (600 maps) of scenarios. The experiments conducted using the FlightGear simulator showed that all methods could land the aircraft appropriately for about 67\% of scenarios considered. The type of landing executed by the UAV was evaluated under two situations. First, The UAV landing is evaluated, taking into account the chance to save the UAV without putting a risk on people, properties, or itself. Next, the UAV landing is evaluated, considering the emphasis on saving people and properties without caring about UAV damages. Finally, statistical analysis reveals that combining the greedy heuristic with the genetic algorithms is suitable for this problem. Although this paper deals with path planning, it is not focused on the National Airspace System (NAS). Furthermore, the presence of the ATC is not taken into account.

% Paper 8
% non-segregated airspace - X
% Initial phase of final approach - X
% Complex situations - X
% Bad weather conditions - X
% Minimum Separation - √
% UAS presence - √
% Time as a constraint - √
\par A framework and a formulation for solving path planning problems for multiple heterogeneous UAVs with uncertain service times for each vehicle–target pair is presented by Sundar et al. \cite{sundar2017path}. The vehicles, which differ in their motion constraints and are located at distinct positions at the beginning of the mission, consider a penalty related to the duration of their total service time. The main goal is to find a tour of each vehicle that starts and ends at its respective position. This considers that every target is visited and serviced by some vehicle, and the sum of the total travel distance and the penalty applied to all vehicles are minimized. Furthermore, the authors present the theoretical properties and advantages of using a two-stage stochastic problem formulation to solve this problem instead of using a deterministic expected value formulation. Finally, extensive numerical simulations that compared these two formulations also corroborate the effectiveness of the proposed approach. However, although this research can be adapted to be applied to problems related to the NAS (e.g., aircraft rerouting), this is aimed to be applied to segregated airspace missions. Furthermore, aspects such as bad conditions are not taken into account. Finally, arrival segment constraints are not taken into account.

% Paper 9
% non-segregated airspace - X
% Initial phase of final approach - X
% Complex situations - X
% Bad weather conditions - X
% Minimum Separation - X
% UAS presence - √
% Time as a constraint - √
\par In \cite{frontera2017approximate}, a fast algorithm that finds collision-free 3D paths for small UAS in urban environments is introduced and combined with an algorithm that computes approximate Euclidean shortest paths. This algorithm reduces the number of obstacles present in the pathfinding process,, considering that the studied environments are expressed as three-dimensional scenarios and the objects as vertical polyhedra. The reader should note that this approach aims to reduce the computation time in a more practical situation, i.e., the algorithm proposed is inefficient in complex situations. Thus, there are situations where the algorithm does not perform well, for instance, scenarios that include tall objects. Experimental cases showed that this approach is competitive in terms of speed and solution quality compared to solutions present in the literature for more realistic scenarios. Furthermore, the authors intend to extend this method for more complex scenarios in future works. However, this research does not consider the application in the National Airspace System (NAS). Also, evolutionary approaches (e.g., Particle Swarm Optimization) are not explored as an alternative to solving complex situations. Finally, some important factors impacting aviation for segregated airspace and NAS, such as bad weather conditions, are not considered.

% Paper 10
% non-segregated airspace - X
% Initial phase of final approach - X
% Complex situations - X
% Bad weather conditions - X
% Minimum Separation - X
% UAS presence - √
% Time as a constraint - X
\par In \cite{shakhatreh2017efficient}, an approach that employs a single UAV for providing wireless coverage for indoor users inside a high-rise building under disaster situations (e.g., earthquakes), considering the failures in cellular networks, is proposed. To accomplish this, the authors assume that the locations of indoor users are uniformly distributed on each floor. Furthermore, a Particle Swarm Optimization (PSO) algorithm is used to find an efficient 3D placement of a UAV that minimizes the total transmit power required for the coverage. The experiments, which considered 50 population size, 50 maximum iteration number, and 20 users on each floor, showed that the proposed approach minimizes the total transmit power required to cover the indoor users considering a uniform distribution of users per floor. Note that the authors state that the PSO is chosen due to the problem's intractability, given its characteristics. In conclusion, this research adopts a traditional implementation of the PSO algorithm and adapts it to the problem. However, changes that may improve the time spent finding an appropriate solution in terms of feasibility and fitness, e.g., polarization, are not considered. On the other hand, this effort focus on small UAS, i.e., the NAS is not considered. Finally, complex situations and bad weather conditions are not included in this approach.

% Paper 11
% non-segregated airspace - X
% Initial phase of final approach - X
% Complex situations - √
% Bad weather conditions - X
% Minimum Separation - X
% UAS presence - √
% Time as a constraint - √
\par Jiang et al. \cite{jiang2017method} establish the model of task assignment for UAV in logistics regarding the Vehicle Routing Problems with Time Windows (VRPTW). In the past few years, there has been a growth in research achievements in logistics and UAV separately, whereas the research achievement on the combination of these areas steadies stable. Effective logistics systems and task assignments reduce operating costs and improve transport efficiency. In this context, the model proposed in this research considers different constraints, such as weight coefficients, time-windows constraints, and the constraints of the UAV. Furthermore, the Particle Swarm Optimization (PSO) algorithm is used for solving the task assignment problem due to its suitability for dealing with complex combinatorial optimization problems. Note that the PSO implementation presents some modifications since the original PSO algorithm is only suitable for the continuous space optimization problem. In this paper, the task assignment for UAV is an integer linear programming problem. The conducted experiments showed that the PSO is efficient in solving the problem of task assignment for UAV, and, in comparison with a traditional Genetic Algorithm (GA), this approach presented a higher success rate and a lower average running time.

% Paper 12
% non-segregated airspace - √
% Initial phase of final approach - √
% Complex situations - √
% Bad weather conditions - X
% Minimum Separation - √
% UAS presence - X
% Time as a constraint - X
\par A new optimization problem for solving conflicts is presented by Hong et al. \cite{hong2017nonlinear}. This method allows aircraft to change their heading angle and speed to optimize their trajectory. The performance index is expressed in terms of the variation of the aircraft arrival time caused by conflict resolution maneuvers, i.e., higher performance indices are computed in situations where this time variation is low. In order to accomplish conflict resolution and proper flow management, metering constraints (e.g., aircraft arrival time) are introduced together with separation constraints. In this context, the optimal solution is obtained by utilizing Particle Swarm Optimization (PSO), and numerical, and Monte Carlo simulations are conducted to evaluate the performance of the proposed algorithm. Due to the considerable ease of PSO in solving complex nonlinear problems, several performance indices and constraints are considered without the limitations of linear approximation or a complex procedure, which may involve a certain level of imprecision. The simulation results showed a significant reduction in the variation of the aircraft arrival time and the magnitude of the maneuvers, i.e., heading angle and speed changes.

% Paper 13
% non-segregated airspace - X
% Initial phase of final approach - X
% Complex situations - X
% Bad weather conditions - X
% Minimum Separation - X
% UAS presence - X
% Time as a constraint - √
\par Marinakis et al. \cite{marinakis2017hybrid} deal with the Constrained Shortest Path problem, which is a well-known NP-hard problem, by proposing a new hybridized version of Particle Swarm Optimization (PSO) algorithm, which is a population-based swarm intelligence method, with Variable Neighborhood Search (VNS), which is an algorithm applied to optimize the particles’ position. Although in the proposed algorithm, a different equation for the velocities update of particles is considered, and a new neighborhood topology is employed, an issue of applying the VNS is the identification of the suitable local search method for a given problem. In this sense, a number of continuous local search algorithms are used and tested in a number of modified instances from and further comparisons with classic versions of PSO. Finally, the experiments showed that the proposed algorithm has satisfactory efficiency and results. In future directions, the authors highlight the application of this methodology to more difficult problems.

% Paper 14
% non-segregated airspace - √ 
% Initial phase of final approach - √
% Complex situations - √ 
% Bad weather conditions - X 
% Minimum Separation - √
% UAS presence - X
% Time as a constraint - √
In \cite{yang2016trajectory}, the authors present an optimization algorithm for solving the problem of arrival aircraft trajectory, which aims to find the best solutions for vertical flight profile considering the Required Time of Arrival (RTA) and constraints of Terminal Maneuvering Area (TMA) and aircraft performance. Firstly, the Base of Aircraft Data (BADA), which is an open-source database of aircraft performance aspects and is used in simulation tools, is used to identify the aircraft's aerodynamic and fuel consumption. Then, a method for optimizing the trajectories is proposed based on an Improved Particle Swarm Optimization (IPSO), in which particles' inertia decreases from 1 to about 0.5 as long as they get closer to near-optimal solutions, with Successive Quadratic Programming (SQP). During the optimization process, the IPSO is employed in finding a near-optimal solution, and then, the SQP is used to enable a quicker finding of an accurate solution. Furthermore, this approach is compared to standard PSO, which shows that its performance is effective for trajectory optimization problems. However, this proposal is not focused on aspects of UAS integration during the optimization process. Finally, although fixed weights are employed in our proposal, the parallel architecture considered tends to improve the algorithm efficiency.

% Paper 15
% non-segregated airspace - √
% Initial phase of final approach - X
% Complex situations - √
% Bad weather conditions - X 
% Minimum Separation - √
% UAS presence - X
% Time as a constraint - X
\par Girish \cite{girish2016efficient} proposes a Hybrid Particle Swarm Optimization-local search (HPSO-LS) algorithm, in a rolling horizon framework, for dealing with the Aircraft Landing Problem (ALP), which consists of the allocation of arriving aircraft to runways and the assignment of a landing time to each aircraft. The main goal of this research is to minimize the penalty costs due to delays in landing times. Note that the rolling horizon framework is used as an online optimization strategy considering a fixed time horizon. The presented results showed that the proposed algorithm effectively solves the problem and compares with existing approaches from the literature (e.g., PSO variants) in scenarios involving up to 500 aircraft and five runways, the RH-HPSO-LS showed to be a more appropriate technique for this problem. In future works, the author intends to improve approaches to reduce the computational time requirements for enabling real-world applications, i.e., real-time applications. However, this effort does not cover the final arrival segments design from the final sector. Also, although the techniques employed are interesting for finding good solutions, the optimization is not built considering aspects of UAS integration in the NAS.

% Paper 16
% non-segregated airspace - √
% Initial phase of final approach - √
% Complex situations - √
% Bad weather conditions - X
% Minimum Separation - √
% UAS presence - X
% Time as a constraint - X
\par Ribeiro et al. \cite{ribeiro2016modeling} propose a framework that integrates performance preferences of landing aircraft in Continuous Descent Arrival (CDA) operations that deals with managing building flight trajectories, which are optimized reduce fuel burn and emissions during the descent/approach phase. This approach is special interesting once the maximization of airspace efficiency and capacity, which needs to be addressed considering local airspace requirements and constraints, is related to the optimization of air traffic trajectories. The authors highlight that the Air Traffic Control (ATC) agent, responsible for conducting the air traffic to specific trajectories, employs a Particle Swarm Optimization (PSO) algorithm to build feasible and safe solutions for arrival sequencing. The results showed that considering the data from Brasilia International Airport (SBBR), the proposed approach enabled 77\% of the air traffic to accomplish their desired time window flying. Finally, as future intentions, the authors aim to deal with en-route trajectory conflicts and capacity constraints. However, this research does not consider some aspects, such as bad weather conditions.

% Paper 17
% non-segregated airspace - √
% Initial phase of final approach - √
% Complex situations - X
% Bad weather conditions - X
% Minimum Separation - √
% UAS presence - X
% Time as a constraint - X
\par The authors in \cite{murata2016optimization} focus on the aircraft landing optimization problem considering both the landing routes and the landing order of aircraft. The main goal is to minimize the occupancy time of the airport, which leads to an increase in airspace efficiency. This approach considers dynamic weather conditions and other aircraft’ landing routes changes. To deal with this problem, the hierarchical evolutionary computation is proposed, which generates candidates for the main landing route of all aircraft. Furthermore, a good combination of landing routes for all aircraft is considered to minimize the occupancy time of the airport. The experiments showed that the proposed strategy generates robust and orderly landing routes. However, the results have only been obtained from one simple grid map, which simulates the flying area of the aircraft and further careful qualifications and justifications (e.g., other maps or a different number of aircraft) represent the future intentions of the authors. Furthermore, our proposal considers complex situations in which feasible solutions in terms of efficiency and, especially, in terms of safety are required. Finally, the UAS integration, which may be an important airspace user in the next years, is considered.

% Paper 18
% non-segregated airspace - X
% Initial phase of final approach - X
% Complex situations - X
% Bad weather conditions - X
% Minimum Separation - X
% UAS presence - √
% Time as a constraint - √
\par Narayan et al. \cite{narayan2009computationally} proposes a novel approach for optimizing 3D flight trajectories considering real-time planning deadlines for small UAS operating in challenging environments, i.e., environments with obstacles. In this approach, which generates feasible solutions, sets of candidate smooth flight trajectories are generated, and, considering that in typical small UAS operations, multiple objectives may exist, a multi-objective optimization is employed since it may allow the discovery of solutions that better reflects overall mission requirements. Note that, in this context, real-time planning constraints may be imposed during the optimization process to avoid obstacles in the immediate path, and this approach considers a novel Computationally Adaptive Trajectory Decision (CATD) optimization system to manage, calculate and schedule parameters associated with trajectories building to ensure that the feasible solutions are offered taking processing duration as a constraint. In conclusion, the authors point out that this approach may potentially be a more efficient use of the computational time available. However, this research is intended to be applied to segregated airspaces. Furthermore, weather conditions are not taken into account.

% Paper 19
% non-segregated airspace - √
% Initial phase of final approach - √
% Complex situations - X
% Bad weather conditions - X
% Minimum Separation - √
% UAS presence - X
% Time as a constraint - √
\par The authors in \cite{ji2014online} propose an online method based on the Estimation of Distribution Algorithm (EDA), which has become a hot topic in the field of evolutionary computing, for the real-time Aircraft Arrival
Sequencing and Scheduling optimization problem. This problem, considered a hot topic in Air Traffic Control (ATC) contributions, has been proven to be an NP-hard problem. Although many efforts have been made by modeling this problem in a static case, the air traffic environment in the airport is dynamic and constantly changing. Since new aircraft are arriving at the airport continually, the corresponding adjustments should be considered for the scheduling definition. In this context, the method focuses on aircraft that have already arrived at the TMA but have not been assigned to land. The experiments highlighted that the method effectively achieves appropriate solutions for the Aircraft Arrival Sequencing and Scheduling optimization problem. However, this contribution does not include the operation of UAS in the NAS and all challenges it brings to the sequencing problem. Furthermore, bad weather conditions are not taken into account. Finally, the fitness evaluation does not consider the impacts of a given sequencing solution on the ATC.

% Paper 20
% non-segregated airspace - √
% Initial phase of final approach - √
% Complex situations - √
% Bad weather conditions - X
% Minimum Separation - √
% UAS presence - X
% Time as a constraint - √
\par Bennell et al. \cite{bennell2017dynamic} deal with scheduling aircraft landings on a single runway. The time window constraints for each aircraft’s landing and the minimum separation between consecutive aircraft and, consequently, consecutive landings are two important metrics for the sequencing problem. Note that the separation between aircraft depends on specific factors, such as the weight classes. Thus, a multi-objective formulation that considers both the runway metrics (throughput, earliness, and lateness) and the fuel cost related to aircraft maneuvers and additional flight time is employed to achieve the landing schedule. This proposal also considers the static/off-line problem, in which details of the arriving flights are provided in advance, and the dynamic/online problem, in which the flight arrival information becomes available over time. The experiments showed that efficient runway throughput results were achieved for both static and dynamic problems, considering the employment of different meta-heuristics.

% Summing up
\par In this section, the works related to air traffic sequencing optimization were presented. Different aspects are covered by each work, but in order to identify the similarities and differences, Table \ref{tab:relatedWorksAirTrafficOptimization} presents all works, which are classified as follows:
\begin{itemize}
    \item \textbf{National Airspace System (NAS):} Indicates if the optimization method is intended to be applied in situations of NAS;
    \item \textbf{Final Arrival Segment Definition (FASD):} Indicates if the proposed method is focused on the final arrival segments design;
    \item \textbf{Complex Situations (CS):} Indicates if the optimization method is developed considering the sequencing of many aircraft, which constitute a complex situation;
    \item \textbf{Bad Weather Conditions (BWC):} Indicates if the proposed solution takes bad weather conditions into account;
    \item \textbf{Minimum Separation (MS):} Indicates if the proposed method applies minimum separations for each aircraft in order to maintain the safety levels;
    \item \textbf{UAS Presence (UAS):} Indicate if the proposed solution considers the presence of the UAS and its impacts on sequencing;
    \item \textbf{Time as a Constraint (TC):} Indicate if processing duration is analyzed in the approach, i.e., if the problem faced is a real-time problem.
\end{itemize}

\begin{table}[]
\centering
\caption{Review of UAS traffic sequencing optimization in the National Airspace System (NAS).}
\label{tab:relatedWorksAirTrafficOptimization}
\begin{tabular}{c|
>{\columncolor[HTML]{009901}}c |
>{\columncolor[HTML]{FD6864}}c |
>{\columncolor[HTML]{009901}}c |
>{\columncolor[HTML]{FD6864}}c |
>{\columncolor[HTML]{009901}}c |
>{\columncolor[HTML]{FD6864}}c |
>{\columncolor[HTML]{009901}}c }
\hline\hline
% \cellcolor[HTML]{9B9B9B}{\color[HTML]{FFFFFF} Related Work} & \cellcolor[HTML]{9B9B9B}{\color[HTML]{FFFFFF} NSA} & \cellcolor[HTML]{9B9B9B}{\color[HTML]{FFFFFF} FASD} & \cellcolor[HTML]{9B9B9B}{\color[HTML]{FFFFFF} CS} & \cellcolor[HTML]{9B9B9B}{\color[HTML]{FFFFFF} BWC} & \cellcolor[HTML]{9B9B9B}{\color[HTML]{FFFFFF} MS} & \cellcolor[HTML]{9B9B9B}{\color[HTML]{FFFFFF} UAS} & \cellcolor[HTML]{9B9B9B}{\color[HTML]{FFFFFF} TC} \\ \hline
\cellcolor[HTML]{FFFFFF}{Related Work} & \cellcolor[HTML]{FFFFFF}{NAS} & \cellcolor[HTML]{FFFFFF}{FASD} & \cellcolor[HTML]{FFFFFF}{CS} & \cellcolor[HTML]{FFFFFF}{BWC} & \cellcolor[HTML]{FFFFFF}{MS} & \cellcolor[HTML]{FFFFFF}{UAS} & \cellcolor[HTML]{FFFFFF}{TC} \\ \hline\hline
\cite{alonso2016multiobjective}                                         & {\color[HTML]{FFFFFF} \checkmark}                           & {\color[HTML]{FFFFFF} X}                          & {\color[HTML]{FFFFFF} \checkmark}                          & {\color[HTML]{FFFFFF} X}                           & {\color[HTML]{FFFFFF} \checkmark}                          & {\color[HTML]{FFFFFF} X}                           & {\color[HTML]{FFFFFF} \checkmark}                          \\ \hline
\cite{gao2012cooperative}                                                 & {\color[HTML]{FFFFFF} \checkmark}                           & {\color[HTML]{FFFFFF} X}                          & {\color[HTML]{FFFFFF} \checkmark}                          & {\color[HTML]{FFFFFF} X}                           & {\color[HTML]{FFFFFF} \checkmark}                          & {\color[HTML]{FFFFFF} X}                           & \cellcolor[HTML]{FD6864}{\color[HTML]{FFFFFF} X}  \\ \hline
\cite{ahmed2017evolutionary}                                            & {\color[HTML]{FFFFFF} \checkmark}                           & \cellcolor[HTML]{009901}{\color[HTML]{FFFFFF} \checkmark}  & {\color[HTML]{FFFFFF} \checkmark}                          & {\color[HTML]{FFFFFF} X}                           & {\color[HTML]{FFFFFF} \checkmark}                          & {\color[HTML]{FFFFFF} X}                           & \cellcolor[HTML]{FD6864}{\color[HTML]{FFFFFF} X}  \\ \hline
\cite{sama2017scheduling}                                                & {\color[HTML]{FFFFFF} \checkmark}                           & {\color[HTML]{FFFFFF} X}                          & {\color[HTML]{FFFFFF} \checkmark}                          & {\color[HTML]{FFFFFF} X}                           & {\color[HTML]{FFFFFF} \checkmark}                          & {\color[HTML]{FFFFFF} X}                           & {\color[HTML]{FFFFFF} \checkmark}                          \\ \hline
\cite{sama2017optimal}                                                & {\color[HTML]{FFFFFF} \checkmark}                           & \cellcolor[HTML]{009901}{\color[HTML]{FFFFFF} \checkmark}  & {\color[HTML]{FFFFFF} \checkmark}                          & {\color[HTML]{FFFFFF} X}                           & {\color[HTML]{FFFFFF} \checkmark}                          & {\color[HTML]{FFFFFF} X}                           & \cellcolor[HTML]{FD6864}{\color[HTML]{FFFFFF} X}  \\ \hline
\cite{sama2017metaheuristics}                                                & {\color[HTML]{FFFFFF} \checkmark}                           & \cellcolor[HTML]{009901}{\color[HTML]{FFFFFF} \checkmark}  & {\color[HTML]{FFFFFF} \checkmark}                          & {\color[HTML]{FFFFFF} X}                           & {\color[HTML]{FFFFFF} \checkmark}                          & {\color[HTML]{FFFFFF} X}                           & {\color[HTML]{FFFFFF} \checkmark}                          \\ \hline
\cite{silva2017heuristic}                                            & \cellcolor[HTML]{FD6864}{\color[HTML]{FFFFFF} X}   & {\color[HTML]{FFFFFF} X}                          & \cellcolor[HTML]{FD6864}{\color[HTML]{FFFFFF} X}  & \cellcolor[HTML]{009901}{\color[HTML]{FFFFFF} \checkmark}   & \cellcolor[HTML]{FD6864}{\color[HTML]{FFFFFF} X}  & \cellcolor[HTML]{009901}{\color[HTML]{FFFFFF} \checkmark}   & {\color[HTML]{FFFFFF} \checkmark}                          \\ \hline
\cite{sundar2017path}                                              & \cellcolor[HTML]{FD6864}{\color[HTML]{FFFFFF} X}   & {\color[HTML]{FFFFFF} X}                          & \cellcolor[HTML]{FD6864}{\color[HTML]{FFFFFF} X}  & {\color[HTML]{FFFFFF} X}                           & {\color[HTML]{FFFFFF} \checkmark}                          & \cellcolor[HTML]{009901}{\color[HTML]{FFFFFF} \checkmark}   & {\color[HTML]{FFFFFF} \checkmark}                          \\ \hline
\cite{frontera2017approximate}                                            & \cellcolor[HTML]{FD6864}{\color[HTML]{FFFFFF} X}   & {\color[HTML]{FFFFFF} X}                          & \cellcolor[HTML]{FD6864}{\color[HTML]{FFFFFF} X}  & {\color[HTML]{FFFFFF} X}                           & \cellcolor[HTML]{FD6864}{\color[HTML]{FFFFFF} X}  & \cellcolor[HTML]{009901}{\color[HTML]{FFFFFF} \checkmark}   & {\color[HTML]{FFFFFF} \checkmark}                          \\ \hline
\cite{shakhatreh2017efficient}                                          & \cellcolor[HTML]{FD6864}{\color[HTML]{FFFFFF} X}   & {\color[HTML]{FFFFFF} X}                          & \cellcolor[HTML]{FD6864}{\color[HTML]{FFFFFF} X}  & {\color[HTML]{FFFFFF} X}                           & \cellcolor[HTML]{FD6864}{\color[HTML]{FFFFFF} X}  & \cellcolor[HTML]{009901}{\color[HTML]{FFFFFF} \checkmark}   & \cellcolor[HTML]{FD6864}{\color[HTML]{FFFFFF} X}  \\ \hline
\cite{jiang2017method}                                            & \cellcolor[HTML]{FD6864}{\color[HTML]{FFFFFF} X}   & {\color[HTML]{FFFFFF} X}                          & {\color[HTML]{FFFFFF} \checkmark}                          & {\color[HTML]{FFFFFF} X}                           & \cellcolor[HTML]{FD6864}{\color[HTML]{FFFFFF} X}  & \cellcolor[HTML]{009901}{\color[HTML]{FFFFFF} \checkmark}   & {\color[HTML]{FFFFFF} \checkmark}                          \\ \hline
\cite{hong2017nonlinear}                                                & {\color[HTML]{FFFFFF} \checkmark}                           & \cellcolor[HTML]{009901}{\color[HTML]{FFFFFF} \checkmark}  & {\color[HTML]{FFFFFF} \checkmark}                          & {\color[HTML]{FFFFFF} X}                           & {\color[HTML]{FFFFFF} \checkmark}                          & {\color[HTML]{FFFFFF} X}                           & \cellcolor[HTML]{FD6864}{\color[HTML]{FFFFFF} X}  \\ \hline
\cite{marinakis2017hybrid}                                           & \cellcolor[HTML]{FD6864}{\color[HTML]{FFFFFF} X}   & {\color[HTML]{FFFFFF} X}                          & \cellcolor[HTML]{FD6864}{\color[HTML]{FFFFFF} X}  & {\color[HTML]{FFFFFF} X}                           & \cellcolor[HTML]{FD6864}{\color[HTML]{FFFFFF} X}  & {\color[HTML]{FFFFFF} X}                           & {\color[HTML]{FFFFFF} \checkmark}                          \\ \hline
\cite{yang2016trajectory}                                                & {\color[HTML]{FFFFFF} \checkmark}                           & \cellcolor[HTML]{009901}{\color[HTML]{FFFFFF} \checkmark}  & {\color[HTML]{FFFFFF} \checkmark}                          & {\color[HTML]{FFFFFF} X}                           & {\color[HTML]{FFFFFF} \checkmark}                          & {\color[HTML]{FFFFFF} X}                           & {\color[HTML]{FFFFFF} \checkmark}                          \\ \hline
\cite{girish2016efficient}                                              & {\color[HTML]{FFFFFF} \checkmark}                           & {\color[HTML]{FFFFFF} X}                          & {\color[HTML]{FFFFFF} \checkmark}                          & {\color[HTML]{FFFFFF} X}                           & {\color[HTML]{FFFFFF} \checkmark}                          & {\color[HTML]{FFFFFF} X}                           & \cellcolor[HTML]{FD6864}{\color[HTML]{FFFFFF} X}  \\ \hline
\cite{ribeiro2016modeling}                                             & {\color[HTML]{FFFFFF} \checkmark}                           & \cellcolor[HTML]{009901}{\color[HTML]{FFFFFF} \checkmark}  & {\color[HTML]{FFFFFF} \checkmark}                          & {\color[HTML]{FFFFFF} X}                           & {\color[HTML]{FFFFFF} \checkmark}                          & {\color[HTML]{FFFFFF} X}                           & \cellcolor[HTML]{FD6864}{\color[HTML]{FFFFFF} X}  \\ \hline
\cite{murata2016optimization}                                              & {\color[HTML]{FFFFFF} \checkmark}                           & \cellcolor[HTML]{009901}{\color[HTML]{FFFFFF} \checkmark}  & \cellcolor[HTML]{FD6864}{\color[HTML]{FFFFFF} X}  & {\color[HTML]{FFFFFF} X}                           & {\color[HTML]{FFFFFF} \checkmark}                          & {\color[HTML]{FFFFFF} X}                           & \cellcolor[HTML]{FD6864}{\color[HTML]{FFFFFF} X}  \\ \hline
\cite{narayan2009computationally}                                             & \cellcolor[HTML]{FD6864}{\color[HTML]{FFFFFF} X}   & {\color[HTML]{FFFFFF} X}                          & \cellcolor[HTML]{FD6864}{\color[HTML]{FFFFFF} X}  & {\color[HTML]{FFFFFF} X}                           & \cellcolor[HTML]{FD6864}{\color[HTML]{FFFFFF} X}  & \cellcolor[HTML]{009901}{\color[HTML]{FFFFFF} \checkmark}   & {\color[HTML]{FFFFFF} \checkmark}                          \\ \hline
\cite{ji2014online}                                                  & {\color[HTML]{FFFFFF} \checkmark}                           & \cellcolor[HTML]{009901}{\color[HTML]{FFFFFF} \checkmark}  & \cellcolor[HTML]{FD6864}{\color[HTML]{FFFFFF} X}  & {\color[HTML]{FFFFFF} X}                           & {\color[HTML]{FFFFFF} \checkmark}                          & {\color[HTML]{FFFFFF} X}                           & {\color[HTML]{FFFFFF} \checkmark}                          \\ \hline
\cite{bennell2017dynamic}                                             & {\color[HTML]{FFFFFF} \checkmark}                           & \cellcolor[HTML]{009901}{\color[HTML]{FFFFFF} \checkmark}  & {\color[HTML]{FFFFFF} \checkmark}                          & {\color[HTML]{FFFFFF} X}                           & {\color[HTML]{FFFFFF} \checkmark}                          & {\color[HTML]{FFFFFF} X}                           & {\color[HTML]{FFFFFF} \checkmark}                          \\ \hline
\end{tabular}
\end{table}

\par This table shows that many related works consider the National Airspace System (NAS), minimum separation, and complex situations. However, only one of them considers bad weather conditions. Note that all works that consider the UAS presence do not integrate them into the NAS, and consequently, all works that deal with NAS do not include the UAS.

% \par Considering that sequencing the air traffic during the final approach can be considered an NP-hard problem \cite{ji2014online}, our novel approach deals with the integration of UAS into the non-segregated airspace. The focus in this research is on the final approach phase considering complex situations. Thus, bad weather conditions are considered and represented as cumulonimbus (CB). Furthermore, a constraint taken into account in the proposed optimization method is the processing duration. This is due to the fact that in these critical situations, the ATCo needs quick responses. Finally, the optimization is achieved by considering sequencing duration, which is related to efficiency and represented by the time spent in delivering a set of aircraft to its destination, and ATCo workload, which is related to safety and is represented by the activities performed by ATCo when establishing vectors for each aircraft in order to deliver them safely.

\section{Open challenges}
\label{sec:open_challenges}
\par
\begin{figure}[H]
\centering
\includegraphics[width=1\linewidth]{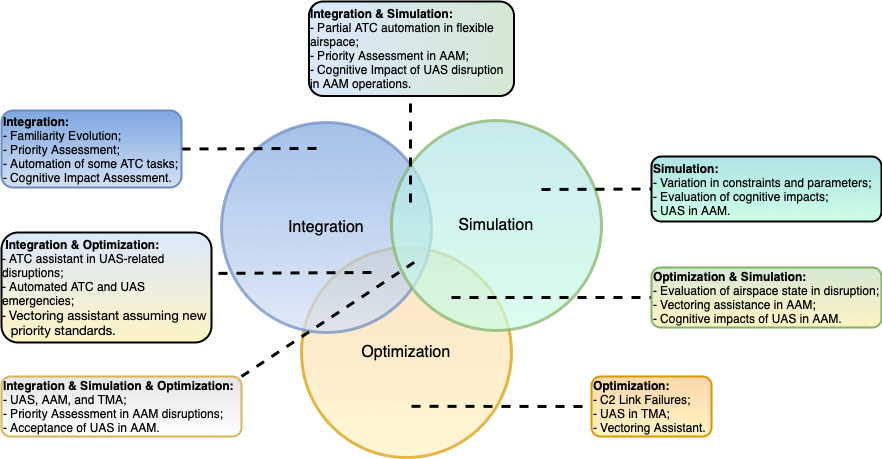}
\caption{Open challenges in UAS Integration, Simulation, Optimization, and their intersections.}
\label{fig:open_challenges}
\end{figure}

 Although this work aims to address specific topics regarding the UAS operation, there are many possibilities for the extension of this effort. Figure \ref{fig:open_challenges} depicts several open challenges in UAS integration, simulation, optimization, and their intersections. In each category, several research directions are identified and described in detail in this Section.

\subsection{UAS Integration}
\begin{itemize}
\item \textbf{Measuring the familiarity evolution of different aircraft types throughout the years:} An open challenge in the UAS context refers to the measurement of the familiarity evolution of different aircraft types (e.g., UAS) as it is dependent on several human factors and social acceptance. Although some initiatives have started to investigate this aspect \cite{vu2014impact} \cite{baum2019mindset}, several other directions require further investigation;
\item \textbf{Priority Establishment of UAS sequencing in the National Airspace System (NAS):} Rather than controlling aircraft from the level of familiarity, an alternative approach is to consider the priority levels established and assessed by a standardized scale. One example of prioritized aircraft, nowadays, is the emergency aircraft \cite{wang2022multi} \cite{malakis2010managing} \cite{haghighi2022performance}. Furthermore, a challenge is to identify the UAS priority following the priority list assigned to different aircraft nowadays;
\item \textbf{Automation of some Air Traffic Control (ATC) tasks:} A complex open challenge is the automation of some ATC tasks. For example, approaches such as ATC Maturity Level (AML), which represents the level of maturity and autonomy of a system in terms of acting in controlling manned and unmanned aircraft (e.g., different approaches for modeling the relationship between the autonomous ATC with UAS and between the autonomous ATC and MA can be developed) \cite{ren2022route} \cite{migliorini2022degraded} \cite{bongo2022effect};

\item \textbf{Cognitive impact assessment of UAS integration when emergencies are declared:} Emergencies in the airspace are critical events that need to be carefully managed \cite{warren2015enabling}. As a result, solutions to deal with these events considering the UAS presence is a vital open challenge \cite{mejias2013multi} \cite{ten2017emergency} \cite{lusk2019safe2ditch}.

\end{itemize}
\subsection{UAS Simulation}
\begin{itemize}
\item \textbf{Applying variations in airspace constraints and parameters:} Another future direction is the development of flexible configurations, e.g., variable CB sizes and shapes, CB movements, and changes in the minimum separation of the aircraft depending on their types and on the characteristics of the airspace (e.g., complexity) \cite{pelegrin2022aircraft} \cite{neto2021trajectory};

\item \textbf{Evaluation of cognitive impacts of different aircraft types:} In the roadmap to define priorities, the evaluation of cognitive impacts on ATCos is a pivotal aspect to consider. In fact, this is an evolving area that can significantly change in the next decades \cite{xie2021explanation} \cite{teutsch2022communication}. Thereupon, the design standard procedures relies on human-centered investigations \cite{sangermano2022safety} \cite{pang2021framework};

\item \textbf{Evaluation of UAS integration in the Advanced Aerial Mobility (AAM):} In the next decades, a new layer of the transportation system is planned to be deployed and widely used. Advanced Aerial Mobility (AAM) relies on Electric Vertical Takeoff, and Landing (eVTOL) vehicles \cite{wang2021trajectory} \cite{neto2021trajectory}. Furthermore, the integration of autonomous vehicles in this new environment is also a challenge to be faced in order to ensure future operations are safe and efficient \cite{de2021comparing} \cite{thompson2022survey} \cite{khan2022generic};

\end{itemize}
\subsection{UAS Optmization}
\begin{itemize}
 \item \textbf{Arrival Segment Design Considering Failures in C2 Link:} the C2 link enables the communication between remote pilots and the aircraft \cite{neto2017airspace} \cite{lee2020lost}. According to the contingency operations proposed by ICAO, considering a failure in the communication within the final sector, conducting all aircraft considering the presence of an independent aircraft is complex. Thus, one open challenge relies on how the set of aircraft can be conducted throughout the landing procedure in these situations;

    \item \textbf{Optimization of the UAS operation in the TMA:} there are several situations faced in larger scenarios from the airspace operation perspective that can be considered. For example, the challenge of dealing with several autonomous aircraft. Examples of open challenges are airspace resilience (e.g., in case of problems in airports) \cite{schimpf2022communication} \cite{de2021decentralized} \cite{clark2018resilience} and impacts of weather conditions in a long period of time (e.g., decades)\cite{sridhar1998airspace} \cite{krozel2003future} \cite{huang2022high}. The main idea is to extend the research conducted in the final sector to a larger and more complex area, the TMA;
    
    \item \textbf{Development of Vectoring assistant to reduce impact on workload:} Vectoring assistance is a key features in advanced ATC \cite{westin2020building} \cite{inoue2012cognitive} \cite{wing2022digital}. Although there are some initiatives under development nowadays, it is important to include human aspects in those systems when the UAS is part of the operation \cite{kirwan2001identification} \cite{prevot2000efficient};
\end{itemize}

\subsection{UAS Integration \& UAS Simulation}
    \begin{itemize}
        \item \textbf{Automation of some ATC tasks for flexible airspace configurations:} In cases of flexible airspace configuration (e.g., minimum separation, priorities, flight rules, and disruption), interoperability is essential for ATC assistants. Examples of such scenarios include abnormal operations with C2Link failures \cite{tomic2022acas}, airport (or vertport in AAM) closure \cite{pejovic2009tentative} \cite{lin2018fast}, and AAM operations with aircraft of diverse aerodynamic capabilities (e.g., speed and turning rate) \cite{johnson2022nasa} \cite{afonso2021design} \cite{warren2019effects};
        
        \item \textbf{Priority assessment for different aircraft types in AAM:} Another challenge relies on the definition of different priorities in AAM. The integration of new vehicles (e.g., UAS) hardens the prioritization process due to the lack of operational history \cite{kamienski2015atc} \cite{xiangmin2020survey}. Thereupon, an assessment of aircraft types and different scenarios is needed to establish standard priorities \cite{kopardekar2017safely} \cite{jover2021tactical};
        
        \item \textbf{Cognitive impact assessment of UAS integration in AAM when emergencies are declared:} A simulation effort is also needed in the integration of UAS in AAM operation \cite{garrow2021urban} \cite{neto2021trajectory}. Also, it is pivotal that future directions consider the analysis of emergencies and abnormal AAM operations, including the UAS.
        
    \end{itemize}
    
\subsection{UAS Integration \& UAS Optimization}
    \begin{itemize}
        \item \textbf{ATC assistant for UAS-related disruptions:} ATC automated support is important in several areas of the airspace \cite{trapsilawati2021integration} \cite{lin2021spoken}. Consequently, the integration of UAS requires new capabilities from these systems. The development of ATC supporting systems for UAS-related disruption is another important open challenge;
        
        \item \textbf{Cognitive impact of UAS emergencies considering automated ATC task:} Although some ATC tasks can be automated, the presence of ATCos is paramount \cite{lin2021improving} \cite{sanford1995tailoring}. In this sense, an investigation of how technology and human interaction in unusual scenarios (e.g., UAS emergencies) \cite{erzberger2004transforming} \cite{pongsakornsathien2020human} is in the scope of future works;
        
        \item \textbf{Development of Vectoring assistant based on new priority standards:} Although ATC supporting tools are under development, new constraints pose the need for adaptable systems for future UAS operations \cite{dalamagkidis2008current} \cite{rattanagraikanakorn2018characterizing}. In fact, new vectoring assistance strategies need to be developed based on future airspace priority standards.
    \end{itemize}
    
\subsection{UAS Simulation \& UAS Optimization}
    \begin{itemize}
        \item \textbf{Evaluation of different airspace configurations in disruption:} In case of abnormal events (e.g., emergencies), it is important to understand how the operation can be optimized \cite{serhan2019dynamic}. Hence, the evaluation and optimization of multiple strategies to deal with complex UAS conditions is another open research challenge.
        
        \item \textbf{Development of Vectoring assistant in the AAM context:} Automation of ATC tasks is challenge for several reasons \cite{scheff2020human} \cite{markpaper2020}. For example, standards are under development, and the daily ATC operation is currently being designed. In this sense, the development of approaches to deal with possible ATC configurations and including the UAS is a vital future direction to support safe and efficient AAM operations;
        
        \item \textbf{Evaluation of cognitive impacts of different aircraft types in AAM:} Similarly to the ATC operation, AAM is expected to have aircraft of multiple capabilities (e.g., speed) \cite{9447255}. Considering the UAS presence is critical since the cognitive impacts on human stakeholders (e.g., ACTos and pilots) can be significant \cite{6697830} \cite{6468022}. Thus, this evaluation fosters the maturity evolution of UAS operations in AAM;

    \end{itemize}

\subsection{UAS Integration \& UAS Simulation \& UAS} 
    \begin{itemize}
        \item \textbf{Automation of UAS-enabled AAM control and its interaction with the TMA:} The diverse environment created by the substantial increase of aircraft in the urban space \cite{straubinger2020overview} \cite{thipphavong2018urban} \cite{al2020factors} will require solutions to optimize the interoperability of the airspace. Solutions that assume the UAS are also required for safe and efficient future operations;

        \item \textbf{Priority assessment in AAM disruptions:} AAM is expected to bring various aircraft to fly simultaneously in the urban environment. Abnormal conditions can lead to unsafe states and compromise the system performance \cite{8453455}. Thereupon, it is important to have strategies and standards established for normal and abnormal operations, considering that these priorities can be flexible depending on several factors (e.g., UAS presence) \cite{sridhar2006initial} \cite{mihetec2013utilization};
        
        \item \textbf{Social acceptance evolution of UAS integration in AAM operations:} UAS is a disruptive technology to be present in airspace. Consequently, there is a lack of social acceptance of such aircraft in the airspace (e.g., AAM) \cite{baum2019mindset} \cite{mersha2020towards}. Investigations of how this problem can be mitigated and the most relevant factors to be addressed represent an open challenge in the UAS context.
        
    \end{itemize}

\section{Conclusion}
\label{sec:conclusion}
\par This research presented a comprehensive review of the advancements in the integration of Unmanned Aircraft Systems (UAS) in the National Airspace System (NAS) from different perspectives. These contributions include the presence of UAS in simulation, the final approach, and the optimization of problems related to the interoperability of such systems in the airspace.
Besides, we also highlighted several open challenges and future directions based on the contributions analyzed. Finally, we emphasize the benefits that UAS will bring to society in the next years and reinforce the need for new strategies to deal with the challenges described in this research.

%Bibliography
\bibliographystyle{unsrt}  
\bibliography{bibliography}

\end{document}